\documentclass[12pt]{article}
\usepackage{amsmath,amsthm}
\usepackage{psfrag}
\usepackage{graphicx}
\usepackage{amssymb,latexsym}

\newtheorem{theorem}{Theorem}[section]
\newtheorem{lemma}[theorem]{Lemma}

\newtheorem{corollary}[theorem]{Corollary}

\theoremstyle{remark}
\newtheorem{remark}[theorem]{Remark}

{\hspace*{\fill}$\rule{.3\baselineskip}{.35\baselineskip}$\end{trivlist}}

\newcommand{\bi}{\begin{itemize}}
\newcommand{\ei}{\end{itemize}}
\newcommand{\bt}{\begin{theorem}}
\newcommand{\et}{\end{theorem}}
\newcommand{\bp}{\begin{proof}}
\newcommand{\ep}{\end{proof}}
\newcommand{\be}{\begin{equation}}
\newcommand{\ee}{\end{equation}}
\newcommand{\ben}{\begin{enumerate}}
\newcommand{\een}{\end{enumerate}}

\newcommand{\C}{\mathbb C}

\newcommand{\N}{\mathbb N}
\newcommand{\Z}{\mathbb Z}
\newcommand{\R}{\mathbb R}

\newcommand{\Rscr}{\mathcal R}

\newcommand{\hf}{\frac{1}{2}}

\newcommand{\e}{\varepsilon}
\renewcommand{\th}{\theta}
\newcommand{\m}{\mu}

\renewcommand{\r}{\rho}

\renewcommand{\th}{\theta}
\newcommand{\G}{\Gamma}
\renewcommand{\O}{\Omega}
\renewcommand{\o}{\omega}

\renewcommand{\b}{\beta}

\newcommand{\z}{\zeta}
\newcommand{\x}{\xi}

\renewcommand{\a}{\alpha}
\renewcommand{\l}{\lambda}
\newcommand{\g}{\gamma}
\renewcommand{\d}{\delta}

\newcommand{\Res}{\text{Res }}
\renewcommand{\and}{\text{~~ and ~~}}
\renewcommand{\part}{\partial}
\newcommand{\ra}{\rightarrow}

\DeclareMathOperator{\Col}{Col}
\renewcommand{\Col}{{\rm Col}}
\newcommand{\gt}{\hat \gamma}

\newcommand{\iet}{{i\over 2\e}}

\newcommand{\pt}{{\pi\over 2}}

\newcommand{\mt}{{\m\over 2}}

\newcommand{\smf}{\sqrt{{\m^2\over 4}-1}}

\newcommand{\ipt}{{i\pi\over 2}}

\newcommand{\teta}{\tilde \eta}

\newcommand{\Si}{\Sigma}

\begin{document}

\title{Nonlinear steepest descent  asymptotics for semiclassical limit of integrable systems:
Continuation in the parameter space}
%of matrix Riemann-Hilbert Problems with analytic jump matrices }

\author{Alexander Tovbis\footnote{
Department of Mathematics,
University of Central Florida,
Orlando, FL 32816, email: atovbis@pegasus.cc.ucf.edu~~~Supported by  NSF grant DMS 0508779}  \ and
Stephanos Venakides\footnote{
Department of Mathematics,
Duke University,
Durham, NC 27708, e-mail:
ven@math.duke.edu~~~
Supported by   NSF grant DMS
0707488}}

\begin{abstract}
The initial value problem of an integrable system, such as  the  Nonlinear Schr\" odinger 
equation, is solved by subjecting  the linear eigenvalue problem arising from its Lax pair to inverse scattering,  
and, thus, transforming it to a matrix Riemann-Hilbert problem (RHP) in the spectral variable. 
In the semiclassical limit, the 
method of nonlinear steepest descent (\cite{DZ1}, \cite{DZ2}), 
supplemented by the  $g$-function mechanism (\cite{DVZ2}), is applied to this 
RHP to produce explicit asymptotic 
solution formulae for the integrable system. These formule
are based on a hyperelliptic Riemann 
surface $\Rscr=\Rscr(x,t)$ in the spectral variable, where
the space-time variables 
$(x,t)$ play the role of external parameters. The curves in the $x,t$ plane, separating regions of different genuses
of $\Rscr(x,t)$, are called breaking curves or
nonlinear caustics.  The genus of $\Rscr(x,t)$ is related to 
 the number of oscillatory phases in the 
asymptotic solution of the integrable system at the point $x,t$. 
An evolution theorem  (\cite{TVZ1}) guarantees the continuous evolution of the 
asymptotic solution in space-time away from the breaking curves.

In the case of the analytic scattering data $f(z;x,t)$
(in the NLS case, $f$ is a  normalized logarithm of the reflection coefficient with time evolution included), the primary role in the breaking mechanism is played by a phase
function $\Im h(z;x,t)$, which is  closely related to the $g$ function. Namely, a break
can be caused (\cite{TVZ1})  either through the change
of topology of zero level curves of  $\Im h(z;x,t)$ (regular break), 
or through the interaction of zero level curves of  $\Im h(z;x,t)$ with 
singularities of $f$ (singular break). Every time a breaking curve in the $x,t$ 
plane is reached, one has to prove the validity of the nonlinear steepest descent asymptotics 
in a region across the curve.

In this paper we prove 
that in the case of a regular break, the nonlinear steepest descent asymptotics can be  ``automatically'' continued through the breaking curve
%was caused by the changing topology of zero level curves of of $\Im h(z;x,t)$.
(however, the expressions for the asymptotic solution will be different 
on the different sides of the curve).
%in the regions of different genuses). 
Our proof is based on the determinantal formula for $h(z;x,t)$ and its 
space and time derivatives, obtained in \cite{TV1}, \cite{TV2}.
Although the results are stated and proven for the  focusing NLS equation, it is clear 
(\cite{TV2})  that
they can be  reformulated for AKNS systems, as well as for the nonlinear steepest descend method
in  a more general setting. 

\end{abstract}
\maketitle

\section{Introduction}\label{body}

The nonlinear steepest descent method, 
%of Deift and Zhou, 
introduced in \cite{DZ1}, \cite{DZ2}, and its extension through the
$g$-function mechanism  introduced in \cite{DVZ2}, is widely used
for asymptotic analysis of matrix Riemann-Hilbert problems (RHPs) with analytic jump matrices (that depend 
on additional parameters).  Remarkable   
recent success stories of 
%using the nonlinear steepest descent 
this method  in such diverse areas as integrable 
systems, orthogonal polynomials, random matrices, approximation theory, etc., can be found, for example, in
\cite{Deiftconf}.
Let one of the additional  parameters in the jump matrix, we denote it $\e$, be a  small
(semiclassical) parameter of the RHP. All the other parameters  are called external parameters; particular external 
parameters considered 
in this paper are $x,t$, which have the meaning of space and time variables for the NLS equation. 
The $g$-function mechanism, when applicable, can be viewed as a way of calculating the 
leading order term of the $\e$ asymptotics to the solution of a matrix RHP;
it consists of reducing the matrix RHP to a scalar, independent of $\e$ (but dependent on $x,t$) RHP \eqref{rhpg}
for the unknown function $g(z)=g(z;x,t)$, which is also a subject of additional requirements: modulation equations  \eqref{modeq} and sign distributions  \eqref{ineq}. There is an underlying hyperelliptic Riemann surface 
$\Rscr=\Rscr(x,t)$, associated 
with $g(z;x,t)$; by the genus of $g(z;x,t)$, as well as the genus of the corresponding matrix RHP,
we understand the genus of  $\Rscr(x,t)$. The genus of $g(z;x,t)$, in general, depends on  external parameters  $x,t$; a point $x,t$, where the genus of $g$ undergoes a change, is
called a breaking point. A curve consisting of breaking points is called {\it breaking curve}  or {\it nonlinear caustics}. Conditions \eqref{modeq}-\eqref{ineq} with a certain genus $N$, which are valid on one side of the 
breaking curve, give no apriori guarantee that  the same conditions with a new value of the genus will be valid on the other side. 
%That means that, speaking generally, conditions \eqref{modeq}-\eqref{ineq} and, 
In particular,
sign distributions \eqref{ineq} 
%(which is the most difficult part)
have to be established anew each time the breaking curve is crossed. For example, it took a lot
of efforts  to prove the  transition from the genus zero to the genus two region, see  Sect. 6.2 of \cite{TVZ1} and the corresponding part of \cite{KMM}.  
Roughly speaking, the key result of the present paper is that sign distributions \eqref{ineq} with the properly chosen genus can be 
{\it automatically exteneded across a breaking curve}, provided that the change of genus (break) is {\it regular},
i.e., that the jump function of the scalar RHP \eqref{rhpg} is analytic on the contour of this RHP, see details below.

The results of this paper are formulated for our model example, which is the matrix RHP that solves the inverse scattering problem for the focusing NLS
\begin{equation}
\label{NLS}
i\epsilon q_t+(\epsilon^2/2) q_{xx}+|q|^2q=0
\end{equation}
with decaying initial data $q(x,0;\e)$ in the semiclassical limit $\e\ra 0$. The contour and the 
jump matrix of this RHP and, accordingly, the contour and the jump function of the corresponding
scalar RHP for $g$, are Schwarz symmetrical (see, for example, \cite{TVZ1}, Sect. 2,1, 2.4). However,
it is an easy observation that our results do not depend on this symmetry and
are applicable in a generic situation, for example, to the semiclassical limit of  AKNS systems.
A more detailed description of $g$-function is given below.

\begin{figure}
%\begin{sideways}
\centerline{
\includegraphics[height=6cm]{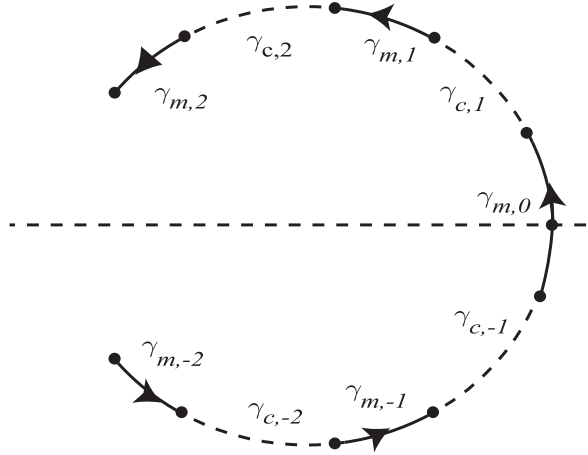}}
%\vskip -.5truecm
\caption{Main and complementary arcs with $n=2$.}
%\hspace{.5cm}
%\psfig{file=../Figures/fig1b.ps}
%\includegraphics[height=9cm]{../Figures/fig1bl.ps}
%\end{center}
%\end{sideways}
\label{mainandcomp}
\end{figure}

Let $\g$ be a Schwarz-symmetrical oriented contour in $\C$ and $f_0(z)$ 
be a  Schwarz-symmetrical analytic  function in some  domain of $ \C$.  We allow
 $f_0(z)$ to have a purely imaginary jump on the real axis. For simplicity, we assume 
$\g$ to be a simple, smooth (except for a finitely many points) contour without self-intersections; moreover, we assume that $\g\cap \R$ consists of one and only one point $\m$.
Let $\g$ consists of $2n+1$, $n\in\N$ , {\em main arcs} $\g_{m,j}$, $j=-n,-n-1,\cdots, n-1,n$, interlaced with $2n$  {\em complementary arcs} $\g_{c,j}$, $j=\pm 1,\pm 2, \cdots,\pm n$, 
see Figure $\ref{mainandcomp}$, and let $\m\in \g_{m,0}$. 
%The union $\g_m$ of all main arcs and the union $\g_c$ of all complementary arcs are also Schwarz symmetrical and 
The main arcs can be considered as branchcuts of  a hyperlliptic Riemann surface $\Rscr$ of genus $N=2n$ that lies at the core of the problem. The endpoints of main arcs are called branchpoints. Branchpoints located in the upper half-plane are denoted $\a_0,\a_2,
\cdots,\a_{4n}$ respectively as we traverse $\g$ in the direction of its orientation.

Because of the  Schwarz symmetry of the problem, main arcs $\g_{m,j}$ and $\g_{m,-j}$,
as well as complementary arcs $\g_{c,j}$ and $\g_{c,-j}$, are Schwarz symmetrical (but their orientation is antisymmetrical)
for all the corresponding $j$s. Unless specified otherwise, we use notations $\g_{m,j}$, $\g_{c,j}$ to denote the union of $\g_{m,j}$ and $\g_{m,-j}$ and the union of $\g_{c,j}$ and $\g_{c,-j}$,   together with their orientations, respectively.
It is clear that branchpoints in the lower half-plane are complex conjugates 
of the corresponding branchpoints $\a_{2j}$, $j=0,2,\cdots, 2n$. We denote them 
$\a_{2j+1}=\overline{\a_{2j}}$.

The complex valued scalar $g$-function satisfies the following Riemann-Hilbert jump and  analyticity conditions:
\begin{align}\label{rhpg}
g_++g_-&=f+W_j ~~\mbox{ on the main arc $\g_{m,j}$, $j=0,\cdots, n$} \cr
g_+ - g_-&=\O_j~~\mbox{ on the complementary arc $\g_{c,j}$, $j= 1,\cdots,  n$ }\cr
g(z) &  ~~\mbox{is analytic in ${\bar \C}\setminus \g$, }\cr
\end{align}
%and $g(\infty)$ is real,
where  the function 
\be\label{f}
f(z)=f_0(z)-zx-2tz^2
\ee
 is a given input to the problem and all $W_j$ and $\O_j$ are real constants. 
%Notice that $\g$ continuously depends on $(x,t)$.
Furthermore, the $g$-function is required to have the following behavior at the branchpoints,
\be\label{modeq}
g(z)=O(z-\a_j)^{3\over 2}~ + ~{\rm analytic~ function~ in~ a~ vicinity~ of }~ \a_j,~~~~~~j=0,1,\cdots,2N+1,
\ee
which imposes $2N+2$ constraints, also known as  {\em modulation equations} on the $2N+2$ branchpoints, where
$N=2n$.
All the branchpoints and all the real constants $W_j$ and $\O_j$ are to be determined
(through \eqref{rhpg}-\eqref{modeq}).
The only given data are the number $N+1$ of branchcuts (or the genus $N$ of the Riemann surface $\Rscr$) and the function $f_0(z)=\iet \ln r(z)$, with $x,t$ being the external parameters (space and time).  Here $r(z)$ is the reflection 
coefficient for some initial data of \eqref{NLS}. 

Solution $g(z)$ of the RHP problem \eqref{rhpg}, which also satisfies modulation
equations \eqref{modeq}, is often known as the $g$-function  of the nonlinear steepest descent method 
(in some papers,  derivative $g'(z)$ is called the $g$-function). However, in order for 
the nonlinear steepest descent asymptotics to work (see, for example, \cite{TVZ3}), 
the  phase function $h=2g-f$ should satisfy the following sign distribution inequalities:
\begin{align}\label{ineq}
\Im h<0& ~~\mbox{ on both sides of each main arc $\g_{m,j}$, $j=0,1,\cdots, n$,} \cr
\Im h>0&~~\mbox{ on at least one side of each complementary arc $\g_{c,j}$, $j=1,\cdots, n$.}\cr
%g(\infty) &\in\R . \cr
\end{align}
%(The corresponding inequalities in the lower half-plane follows from \eqref{ineq} due to Schwarz symmetry.) 
These inequalities show that all the main arcs lie on  zero level curves of $\Im h(z)$ and,
unless prevented by singularities of $f_0(z)$, all the complementary arcs could be continuously deformed so that they also lie  on  zero level curves of $\Im h(z)$ (it is possible that parts of
some complementary arcs would lie on $\R$).
As we continuously deform external parameters $x,t$, the branchpoints $\a_j$  move according to
\eqref{modeq}, pulling (deforming)  main and complemenary arcs of the contour $\g=\g(x,t)$
with them. We say that {\it the nonlinear  steepest descent asymptotics is valid for some 
values of $x,t$ if there exists $n\in\N$, such that all the branchpoints $\a_j$ stay away from
$\R\cup \infty$ and
the solution $g(z;x,t)$ of \eqref{rhpg} 
satisfies \eqref{modeq} and \eqref{ineq}}. If the nonlinear  steepest descent asymptotics 
is valid for some  $x,t$, then the expression for the leading order term (as $\e\ra 0$) of the solution
$q(x,t,\e)$ to the NLS \eqref{NLS} at  $x,t$ that corresponds to the scattering data $r(z)$ is given
in \cite{TVZ1}, Main Theorem. 

%%%%from here
Suppose that the nonlinear steepest descent  asymptotics is valid for some particular value of  $x_*,t_*$.
Then, according to the Evolution Theorem (Theorem
3.2) of \cite{TVZ1}, $g(z;x,t)$ with the same genus $N=2n$ satisfies \eqref{modeq} and \eqref{ineq}  in a neighborhood of $x_*,t_*$ of the $x,t$-plane. If  $x,t$ are evolving further (outside this neighborhood)  along some piecewise-smooth curve $\Si$ in the $x,t$-plane, $x_*,t_*\in \Si$, then
it is possible (\cite{TVZ1}, Section 3) that 
an  inequality of \eqref{ineq}   fails at a  
point $x_b,t_b\in \Si$ (breaking point). 
This failure can be caused by one of the following two reasons: a) regular, when a change of the topology of zero level curves of $\Im h(z)=\Im h(z;x,t)$ at $(x,t)=(x_b,t_b)$ affects contour $\g$; 
b) singular, 
when the contour 
$\g=\g(x,t)$ interacts (collides or encircles) with singularities (including branchcuts) of $f_0(z)$ at $(x,t)=(x_b,t_b)$.
 
The goal of this paper is to address the regular breaking (scenario a)), leaving the case of the singular breaking (scenario b)) to be addressed elsewhere. Let the genus of $g(z;x_*,t_*)$ be $N=2n$.
According to \cite{TVZ1}, Section 3, the change of topology
of zero level curves of $\Im h(z)$ at the breaking point $x_b,t_b$ contains two generic possibilities: 
i) two branches of zero level curve of $\Im h(z)$ collide at some point $z_0\in\g$
that is not a branchpoint; 
%such point $z_0$ is called a double point; 
ii) two adjacent branchpoints collide at some point $z_0$ (collision of nonadjacent branchpoints creates a loop that  encircles some singularities). In any case, $z_0$ is called a breaking point in the spectral plane that corresponds
to the breaking point $x_b,t_b$ in the $x,t$ plane.
In the case i) we can plant a pair of branchpoints at the breaking point $z_0$  and another pair of branchpoints
at the conjugated breaking point $\bar z_0$. That allows us to consider the corresponding hyperelliptic surface
$\Rscr=\Rscr(x,t)$ as having genus $N$ at the breaking point $x_b,t_b$ before planting the branchpoints and, simultaneously, as having 
genus $N+2$  after the planting. As we evolve further along $\Si$,
a new pair of main arcs (if $z_0\in\g_c$) or  of complementary arcs (if $z_0\in\g_m$)
with endpoints evolving from $z_0$ and from $\bar z_0$ opens up. 
%Thus, the genus of $\Rscr=\Rscr(x,t)$ changes from $N$ to $N+2$. 
The case ii) can be described  by evolving
along $\Si$ through the breaking point $x_b,t_b$ 
%from the region of genus $N$ to the region of genus $N-2$, i.e., 
in the opposite direction. By removing a pair of colliding branchpoints (and their conjugates), we reduce the genus of $\Rscr$ by two, say, from $N$ to $N-2$. 
%A point $x,t$ on $\Si$ where the change of topology occurs is called a breaking point. The hyperelliptic surface 
%$\Rscr$ at the breaking point has genus $N$. Howver, if we count a pair of arcs (main or complementary),
%degenerated into the points $z_0$ and $\bar z_0$ respectively, the genus will be $N+2$. 
In degenerate cases, several zero level curves of $\Im h(z)$  meet at the same point
$z_0$, which may or may not be a branchpoint. Then 
\be\label{degdeg}
h(z;x_b,t_b)=C+ O(z-z_0)^{m}~,
\ee
where $2m\in\Z^+$ and $C$ is a real constant. $m$ is called the degree of degenerate  breaking point 
$z_0$.
%For example, the degree of a double point  $z_0$  is two.
Note that if the breaking point $z_0$ is also a branchpoint, then $m$ is a half-integer number,
otherwise, $m$ is an integer. The number of zero level curves of $\Im h(z)$, emanating from 
$z_0$, is $2m$, and the number of the branchpoints, ``born'' at the breaking point $z_0$,
is $2m-2$. For example,  two branchpoints emanate from  $z_0$ of degree two (called a double point),
three branchpoints emanate from $z_0$ of degree $5/2$ (called a triple point), etc.  
 In \cite{TVZ1}, the only triple
point was the  point at the tip (corner) of the breaking curve; it was the point where the very first
break (in the process of time evolution) occurs. It is possible that there are several
breaking points in the spectral plane (without counting complex conjugated points) that correspond
to the same breaking point $x_b,t_b$, for example, when several inequalities of \eqref{ineq}
fail at $x_b,t_b$. Such breaking points $x_b,t_b$ are  degenerate (non-generic). It is shown in
Sect. \ref{regcontpr} that degenerate breaking points are isolated points in the $x,t$-plane. 
%that can be avoided by a small deformation of the path.

Let $g^{(N)}(z)$  denote the solution of the RHP \eqref{rhpg} with $N+1=2n+1$ main arcs, i.e., 
$g^{(N)}(z)$ denotes a $g$ function of the genus $N$, and let $h^{(N)}(z)=2g^{(N)}(z)-f(z)$.  
The Degeneracy Theorem (Theorem 3.1)  of \cite{TVZ1} states that 
$h^{(N+2)}(z;x_b,t_b)\equiv h^{(N)}(z;x_b,t_b)$, provided that $x_b,t_b$ is a regular breaking point. 
The Degeneracy Theorem is an important tool in tracking the signs of $\Im h(z;x,t)$, and with them, the validity of 
of the nonlinear steepest descent asymptotics, through 
breaking points. However, it does not guarantee the correct sign distribution, i.e., inequalities
\eqref{ineq}, past the breaking point, i.e., in the genus $N+2$ or in the genus $N-2$ regions.
For example, in the case i) it does not guarantee that the signs of $\Im h$
around the newborn arc  are correct, i.e., that  the corresponding inequality from  \eqref{ineq}
is satisfied  
(signs around all the other arcs are correct by the continuity argument). 
To track the signs of $\Im h(z)$ through the breaking point, it would be very helpful 
to establish that not only $h^{(N+2)}(z)$ and $h^{(N)}(z)$ are equivalent at the breaking point, but that so are their partial derivatives with respect to external parameters, i.e., $h_x$ and $h_t$. The latter statements do not follow from the Degeneracy Theorem directly,
since $h^{(N+2)}(z;x_b,t_b)\equiv h^{(N)}(z;x_b,t_b)$ only at the breaking point $x_b,t_b$, but not  in any vicinity 
of this point.
% on $\Si$. 

The key observation of this  paper is  that, in fact, 
\be\label{hxhtint}
h^{(N+2)}_x(z;x_b,t_b)\equiv h^{(N)}_x(z;x_b,t_b)~~~~{\rm and}~~~~ h^{(N+2)}_t(z;x_b,t_b)\equiv h^{(N)}_t(z;x_b,t_b)
\ee
at any regular and generic breaking point $x_b,t_b$. The proof of  \eqref{hxhtint} involves the determinant formula from \cite{TV2}. 
Equations \eqref{hxhtint} allow us to prove
that the nonlinear steepest descent asymptotics {\it is always preserved} when one passes through a
regular and generic breaking point, provided that the genus of the problem is adjusted accordingly. 
Speaking somewhat  lousely, we can formulate  the following  {\it regular continuation principle}.

{\it \underbar{Regular continuation principle for the nonlinear steepest descent asymptotics:} Let the nonlinear steepest descent asymptotics 
for solution $q(x,t,\e)$ of the NLS \eqref{NLS} 
be valid at some  point $(x_b,t_b)$. If $(x_*,t_*)$ is an arbitrary point, 
connected with $(x_b,t_b)$ by a
piecewise-smooth path $\Sigma$,  if the countour 
$\g(x,t)$ of the RHP \eqref{rhpg} does not interact with  singularities of $f_0(z)$
as $(x,t)$ varies from $(x_b,t_b)$ to   $(x_*,t_*)$ along $\Sigma$, 
and if all the branchpoints are bounded  
and stay away from the real axis,
then  the nonlinear steepest descent asymptotics (with the proper choice of the  genus) is also valid at $(x_*,t_*)$.}

%COMMENT: We need to address the degenerate breaking points issue.
This principle will be proved in Section \ref{regcontpr}.
% under the additional assumption that all the breaking points on $\Si$ between $(x_0,t_0)$ and   $(x_*,t_*)$ are double  points. 
%either double or triple points. 
%The authors, however, believe that the principle is correct as stated above.
Some important  facts about the determinantal formula are provided in Section \ref{detf},
whereas formula  \eqref{hxhtint} is proven in Theorem  \ref{equalh}, Section \ref{conthxtbre}.

\section{Determinantal formula}\label{detf}

Theorem  \ref{equalh},  which is the central part of the regular continuation principle, 
is also an important advancement of the Degeneracy Theorem from \cite{TVZ1}.
Its proof is based on the determinant representation of $h$ and its immediate consequences, obtained 
in \cite{TV1}, \cite{TV2}. Some basic facts from \cite{TV2} are given in this section.

Assuming that $W_0$, $W_j,\O_j$, $j=1,2,\cdots, n$, and $\a_j,~j=0,1,\cdots,4n+1$,
are  known, the solution
to the RHP \eqref{rhpg} is given by
\begin{equation}\label{gform1}
g(z)={{R(z)}\over{2\pi i}}\left[ \int_\g{{f(\z)}\over{(\z-z)R(\z)_+}}d\z
+\sum_{j=0}^n \int_{\g_{m,j}}{{W_j}\over{(\z-z)R(\z)_+}}d\z
+\sum_{j=1}^n\int_{\g_{c,j}}{{\Omega_j}\over{(\z-z)R(\z)}} d\z\right]~.
\end{equation}
where the  radical $R(z)=\sqrt{\prod_{j=0}^{4n+1}(z-\a_j)}$ has branchcuts $\g_{m,j}$, $j=0,1,\cdots,n$, i.e.,
$\Rscr$ is
the Riemann surface (of the genus $N=2n$) of the  radical $R(z)$. We fix the branch of $R(z)$ by the requirement
\be\label{Rlim}
\lim_{z\ra\infty} \frac{R(z)}{z^{N+1}}=-1
\ee
on the main sheet of $\Rscr$.
%the  union ranges over all the main arcs while the  summation
%over main and complementary arcs excludes the $0$th main arc.

\begin{figure}
%\begin{sideways}
\centerline{
\includegraphics[height=6cm]{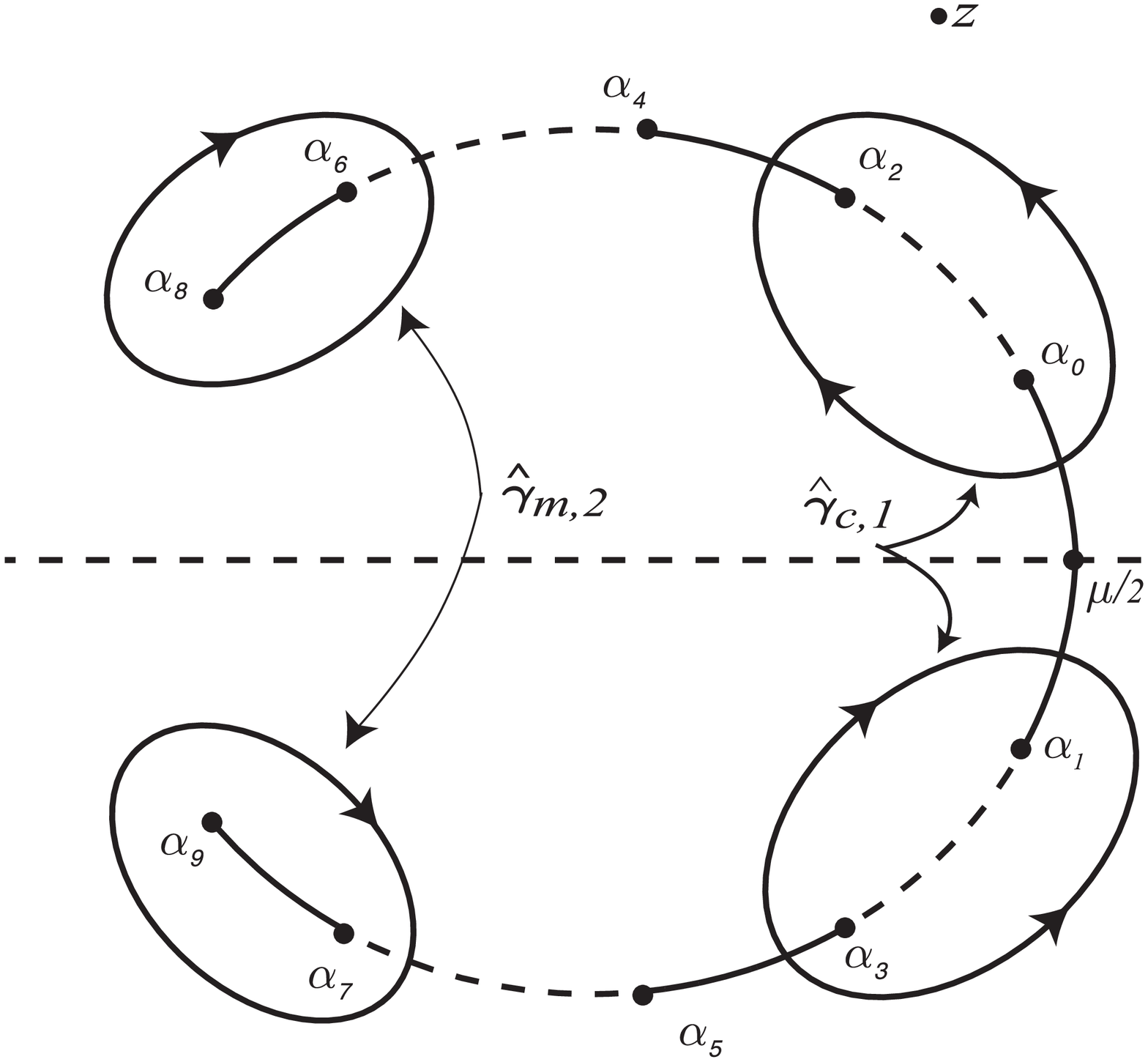}}
%\vskip -.5truecm
\caption{Contours  $\gt_{m,2},~\gt_{c,1}$ }
%\hspace{.5cm}
%\psfig{file=../Figures/fig1b.ps}
%\includegraphics[height=9cm]{../Figures/fig1bl.ps}
%\end{center}
%\end{sideways}
\label{figloop}
\end{figure}

Expressing the integrals over main and complementary arcs as integrals over the loops shown in Fig. \ref{figloop}, i.e., as $\a$ cycles  and as combinations of $\b$ cycles of the hyperelliptic surface $\Rscr$,
we obtain
\begin{equation}\label{gform2}
g(z)={{R(z)}\over{4\pi i}}\left[ \oint_{\gt}{{f(\z)}\over{(\z-z)R(\z)}}d\z
+\sum_{j=0}^n \oint_{\gt_{m,j}}{{W_j}\over{(\z-z)R(\z)}}d\z
+\sum_{j=1}^n\oint_{\gt_{c,j}}{{\Omega_j}\over{(\z-z)R(\z)}} d\z\right],
\end{equation}
where  the loops   $\gt_{m,j}$ around  main arcs $\g_{m,j}$ have negative (clock-wise)
orientation (an $\a$ cycle)
and the loops $\gt_{c,j}$ around complementary arcs $\g_{c,j}$ have   positive
(counterclock-wise) orientation. Here the part of $\gt_{c,j}$ on the main sheet of $\Rscr$
has the same orientation as $\g_{c,j}$ and the part of $\gt_{c,j}$ on the secondary sheet of $\Rscr$
has the opposite orientation (a $\b$ cycle). Alternatively, $\gt_{c,j}$ can be considered
as a union of two arcs on the main sheet of $\Rscr$ surrounding $\g_{c,j}$ with opposite
orientations. The loop $\gt$ is a negatively oriented contour surrounding $\g$. All loops are contained
in $S$ and are
contractible  to their corresponding arcs without passing through $z$ (that mean that the loops are
pinched to their respective contours at the points of nonanalyticity of $f_0(z)$)..

Deforming $\gt$ so that  $z$ {\it becomes inside the loop} $\gt$ and still outside the loops $\gt_{m,j}$ and
$\gt_{c,j}$, we obtain
\begin{equation}\label{hform}
h(z)={{R(z)}\over{2\pi i}}\left[ \oint_{\gt}{{f(\z)}\over{(\z-z)R(\z)}}d\z
+\sum_{j=0}^n \oint_{\gt_{m,j}}{{W_j}\over{(\z-z)R(\z)}}d\z
+\sum_{j=1}^n\oint_{\gt_{c,j}}{{\Omega_j}\over{(\z-z)R(\z)}} d\z\right],
\end{equation}
where
\be \label{h=2g-f}
h(z)=2g(z)-f(z).
\ee
The function $h(z)$ is obtained by multiplying $g(z)$ by a factor of $2$ and the
residue $-f$ being picked up as $z$  cuts through the loop $\gt$.

According to \eqref{Rlim} and \eqref{gform2}, $g(z)\sim O(z^{N})$ as $z\ra\infty$. Without
any loss of generality, we can assume that $W_0=0$ (otherwise, replacing the solution $g(z)$ of \eqref{rhpg}
by $g(z)-\hf W_0$, we add $-W_0$ to jump constant $W_j$ on every main arc $\g_{m,j}$, as well as to
$g(\infty)$, without changing any of the jump constants $\O_j$).
The requirement that $g(z)$ is analytic at $z=\infty$, see \eqref{rhpg}, together with the 
Schwarz symmetry define the system of $N=2n$
real linear equations
\begin{equation}\label{moments}
%\mbox{$k$th  moment condition for $g$}: \ \ \ \ \
\oint_{\gt}{\z^k {f(\z)}\over{R(\z)}}d\z
+\sum_{j=1}^N \oint_{\gt_{m,j}}{{W_j\z^k}\over{R(\z)}}d\z
+\sum_{j=1}^N \oint_{\gt_{c,j}}{{\Omega_j\z^k}\over{R(\z)}}d\z=0,
 \ \ \ \ \
k=0,1,\cdots, N-1,
\end{equation}
for $N$ real
 variables  $W_j,\O_j$, $j=1,2,\cdots,n$.
Let us introduce
\be\label{DNLS}
D=\left|\begin{matrix} \oint_{\gt_{m,1}}\frac{ d\z}{R(\z)} &
\cdots & \oint_{\gt_{m,1}}\frac{\z^{2n-1} d\z}{R(\z)} \cr
%&\Im  \oint_{\gt_{m,1}}\frac{\z^{N} d\z}{R(\z)}\cr\cdots &\cdots & \cdots&
\cdots & \cdots & \cdots \cr
 \oint_{\gt_{m,n}}\frac{ d\z}{R(\z)} &
\cdots & \oint_{\gt_{m,n}}\frac{\z^{2n-1} d\z}{R(\z)}
%&\Im  \oint_{\gt_{m,N}}\frac{\z^{N} d\z}{R(\z)}
\cr
\oint_{\gt_{c,1}}\frac{ d\z}{R(\z)} &
\cdots &\ \oint_{\gt_{c,1}}\frac{\z^{2n-1} d\z}{R(\z)}
%&\Im  \oint_{\gt_{c,1}}\frac{\z^{N} d\z}{R(\z)}
\cr
%\cdots &\cdots & \cdots&
\cdots & \cdots & \cdots \cr
 \oint_{\gt_{c,n}}\frac{ d\z}{R(\z)} &
\cdots & \oint_{\gt_{c,n}}\frac{\z^{N-1} d\z}{R(\z)}
%& \Im  \oint_{\gt_{c,N}}\frac{\z^{N} d\z}{R(\z)}
\cr
%\oint_{\gt_{m,0}}\frac{ d\z}{R(\z)} & \cdots & \oint_{\gt_{m,0}}\frac{\z^{N-1} d\z}{R(\z)} &\overline{ \oint_{\gt_{m,0}}\frac{ d\z}{R(\z)}}
%& \cdots &\overline{ \oint_{\gt_{m,0}}\frac{\z^{N-1} d\z}{R(\z)}}&\Im  \oint_{\gt_{m,0}}\frac{\z^{N} d\z}{R(\z)}\cr
\end{matrix}\right|~
\ee
\medskip
and 
\be\label{KNLS}
K(z)= \frac{1}{2\pi i}\times
\left| \begin{matrix}
  \oint_{\gt_{m,1}}\frac{ d\z}{R(\z)} &
\cdots &  \oint_{\gt_{m,1}}\frac{\z^{2n-1} d\z}{R(\z)} &  \oint_{\gt_{m,1}}\frac{d\z}{(\z-z)R(\z)}\cr
%\cdots &\cdots & \cdots&
\cdots & \cdots & \cdots  & \cdots\cr
  \oint_{\gt_{m,n}}\frac{ d\z}{R(\z)} &
\cdots &  \oint_{\gt_{m,n}}\frac{\z^{2n-1} d\z}{R(\z)}
 &\oint_{\gt_{m,n}}\frac{d\z}{(\z-z)R(\z)}\cr
  \oint_{\gt_{c,1}}\frac{ d\z}{R(\z)} &
\cdots &  \oint_{\gt_{c,1}}\frac{\z^{2n-1} d\z}{R(\z)}
 &\oint_{\gt_{c,1}}\frac{d\z}{(\z-z)R(\z)}\cr
%\cdots &\cdots & \cdots&
\cdots & \cdots & \cdots & \cdots \cr
  \oint_{\gt_{c,n}}\frac{ d\z}{R(\z)} &
\cdots &  \oint_{\gt_{c,n}}\frac{\z^{2n-1} d\z}{R(\z)}
&\oint_{\gt_{c,n}}\frac{d\z}{(\z-z)R(\z)}\cr
%
%\oint_{\gt_{m,0}}\frac{ d\z}{R(\z)} & \cdots & \oint_{\gt_{m,0}}\frac{\z^{N-1} d\z}{R(\z)} &\overline{ \oint_{\gt_{m,0}}\frac{ d\z}{R(\z)}}
%& \cdots &\overline{ \oint_{\gt_{m,0}}\frac{\z^{N-1} d\z}{R(\z)}}&\Im  \oint_{\gt_{m,0}}\frac{\z^{N} d\z}{R(\z)}&\oint_{\gt_{m,0}}\frac{d\z}{(\z-z)R(\z)}\cr
%
\oint_{\gt}\frac{f(\z)d\z}{R(\z)} & \cdots
&   \oint_{\gt}\frac{\z^{2n-1} f(\z)d\z}{R(\z)} &
\oint_{\gt}\frac{f(\z)d\z}{(\z-z)R(\z)}\cr
\end{matrix}\right|~.
\ee
%respectively.
%where $\g_m=\g_{m,1}$ and $\g_c=\g_{c,1}$.
%
Note that $D$ can be reduced to the determinant made of basic holomorphic differentials
of $\Rscr$ (\cite{TV2}) and thus $D\ne 0$. The latter implies 
solvability of \eqref{moments} with any $f(z)=f(z;x,t)$ given by \eqref{f}.
%follows from the known fact that $D\ne 0$.
That allows us to obtain 
\be\label{hK}
h(z)=\frac{R(z)}{D}K(z)~,
\ee
where $z$ is inside the loop $\gt$ but outside all other loops $\gt_{m,j},\gt_{c,j}$.
%That will be our standard assumption about the location of $z$ for the rest of the paper, unless specified otherwise. On the other hand, 
Assumption that $z$ is outside the loop $\gt$ yields 
\be\label{gK}
g(z)=\frac{R(z)}{2D}K(z)~.
\ee
Equation \eqref{hK} allows us (\cite{TV2}) to obtain 
\be\label{dhdxtdK}
\frac{d}{dx}h(z)=\frac{R(z)}{D}\frac{\part}{\part x} K(z),~~~
\frac{d}{dt}h(z)=\frac{R(z)}{D}\frac{\part}{\part t} K(z)~.
\ee
Combining  \eqref{dhdxtdK} with  \eqref{KNLS} and  \eqref{f}, one can easily obtain
\be\label{dKdxn}
\frac{\part}{\part x} K(z)=\left| \begin{matrix}
  \oint_{\gt_{m,1}}\frac{ d\z}{R(\z)} &
\cdots &  \oint_{\gt_{m,1}}\frac{\z^{2n-2} d\z}{R(\z)} &  \oint_{\gt_{m,1}}\frac{d\z}{(\z-z)R(\z)}\cr
%\cdots &\cdots & \cdots&
\cdots & \cdots & \cdots  & \cdots\cr
  \oint_{\gt_{m,n}}\frac{ d\z}{R(\z)} &
\cdots &  \oint_{\gt_{m,n}}\frac{\z^{2n-2} d\z}{R(\z)}
 &\oint_{\gt_{m,n}}\frac{d\z}{(\z-z)R(\z)}\cr
  \oint_{\gt_{c,1}}\frac{ d\z}{R(\z)} &
\cdots &  \oint_{\gt_{c,1}}\frac{\z^{2n-2} d\z}{R(\z)}
 &\oint_{\gt_{c,1}}\frac{d\z}{(\z-z)R(\z)}\cr
%\cdots &\cdots & \cdots&
\cdots & \cdots & \cdots & \cdots \cr
  \oint_{\gt_{c,n}}\frac{ d\z}{R(\z)} &
\cdots &  \oint_{\gt_{c,n}}\frac{\z^{2n-2} d\z}{R(\z)}
&\oint_{\gt_{c,n}}\frac{d\z}{(\z-z)R(\z)}\cr
\end{matrix}\right|
\ee
and
\be\label{dKdtn}
\frac{\part}{\part t} K(z)=-2\left| \begin{matrix}
  \oint_{\gt_{m,1}}\frac{ d\z}{R(\z)} &
\cdots &  \oint_{\gt_{m,1}}\frac{\z^{2n-3} d\z}{R(\z)} &  \oint_{\gt_{m,1}}\frac{d\z}{(\z-z)R(\z)}
&  \oint_{\gt_{m,1}}\frac{\z^{2n-1} d\z}{R(\z)}
\cr
%\cdots &\cdots & \cdots&
\cdots& \cdots & \cdots & \cdots  & \cdots\cr
  \oint_{\gt_{m,n}}\frac{ d\z}{R(\z)} &
\cdots &  \oint_{\gt_{m,n}}\frac{\z^{2n-3} d\z}{R(\z)}
 &\oint_{\gt_{m,n}}\frac{d\z}{(\z-z)R(\z)}&  \oint_{\gt_{m,n}}\frac{\z^{2n-1} d\z}{R(\z)}\cr
  \oint_{\gt_{c,1}}\frac{ d\z}{R(\z)} &
\cdots &  \oint_{\gt_{c,1}}\frac{\z^{2n-3} d\z}{R(\z)}
 &\oint_{\gt_{c,1}}\frac{d\z}{(\z-z)R(\z)}&  \oint_{\gt_{c,1}}\frac{\z^{2n-1} d\z}{R(\z)}\cr
%\cdots &\cdots & \cdots&
\cdots& \cdots & \cdots & \cdots & \cdots \cr
  \oint_{\gt_{c,n}}\frac{ d\z}{R(\z)} &
\cdots &  \oint_{\gt_{c,n}}\frac{\z^{2n-3} d\z}{R(\z)}
&\oint_{\gt_{c,n}}\frac{d\z}{(\z-z)R(\z)}&  \oint_{\gt_{c,n}}\frac{\z^{2n-1} d\z}{R(\z)}\cr
\end{matrix}\right|
+\sum_{j=0}^{4n+1}\a_j\frac{\part}{\part x} K(z)~.
\ee

Equations \eqref{DNLS}-\eqref{dKdtn} can be, in fact, extended
to a more general situation, where $f_0(z)$ and contour $\g$ are not necessarily Schwarz-symmetrical
(this would extend the nonlinear steepest descent mehtod from the NLS to some general AKNS systems). 
In particular (see \cite{TV2}):  
\be\label{D}
D=\left|\begin{matrix} \oint_{\G_{m,1}}\frac{ d\z}{R(\z)} &
\cdots & \oint_{\G_{m,1}}\frac{\z^{N-1} d\z}{R(\z)} &\overline{ \oint_{\G_{m,1}}\frac{ d\z}{R(\z)}}
& \cdots &\overline{ \oint_{\G_{m,1}}\frac{\z^{N-1} d\z}{R(\z)}} \cr
%&\Im  \oint_{\gt_{m,1}}\frac{\z^{N} d\z}{R(\z)}\cr
\cdots &\cdots & \cdots& \cdots & \cdots & \cdots \cr
 \oint_{\G_{m,N}}\frac{ d\z}{R(\z)} &
\cdots & \oint_{\G_{m,N}}\frac{\z^{N-1} d\z}{R(\z)} &\overline{ \oint_{\G_{m,N}}\frac{ d\z}{R(\z)}}
& \cdots &\overline{ \oint_{\G_{m,N}}\frac{\z^{N-1} d\z}{R(\z)}}
%&\Im  \oint_{\gt_{m,N}}\frac{\z^{N} d\z}{R(\z)}
\cr
\oint_{\G_{c,1}}\frac{ d\z}{R(\z)} &
\cdots &\ \oint_{\G_{c,1}}\frac{\z^{N-1} d\z}{R(\z)} & \overline{ \oint_{\G_{c,1}}\frac{ d\z}{R(\z)}}
& \cdots &\overline{ \oint_{\G_{c,1}}\frac{\z^{N-1} d\z}{R(\z)}}
%&\Im  \oint_{\gt_{c,1}}\frac{\z^{N} d\z}{R(\z)}
\cr
\cdots &\cdots & \cdots& \cdots & \cdots & \cdots \cr
 \oint_{\G_{c,N}}\frac{ d\z}{R(\z)} &
\cdots & \oint_{\G_{c,N}}\frac{\z^{N-1} d\z}{R(\z)} &\overline{ \oint_{\G_{c,N}}\frac{ d\z}{R(\z)}}
& \cdots &\overline{ \oint_{\G_{c,N}}\frac{\z^{N-1} d\z}{R(\z)}}
%& \Im  \oint_{\gt_{c,N}}\frac{\z^{N} d\z}{R(\z)}
\cr
%\oint_{\gt_{m,0}}\frac{ d\z}{R(\z)} & \cdots & \oint_{\gt_{m,0}}\frac{\z^{N-1} d\z}{R(\z)} &\overline{ \oint_{\gt_{m,0}}\frac{ d\z}{R(\z)}}
%& \cdots &\overline{ \oint_{\gt_{m,0}}\frac{\z^{N-1} d\z}{R(\z)}}&\Im  \oint_{\gt_{m,0}}\frac{\z^{N} d\z}{R(\z)}\cr
\end{matrix}\right|~
\ee
and 
\be\label{K}
K(z)= \frac{1}{2\pi i}\left| \begin{matrix}
  \oint_{\G_{m,1}}\frac{ d\z}{R(\z)} &
\cdots &  \oint_{\G_{m,1}}\frac{\z^{N-1} d\z}{R(\z)} & \overline{ \oint_{\G_{m,1}}\frac{ d\z}{R(\z)}}
& \cdots & \overline{ \oint_{\G_{m,1}}\frac{\z^{N-1} d\z}{R(\z)}}& \oint_{\G_{m,1}}\frac{d\z}{(\z-z)R(\z)}\cr
\cdots &\cdots & \cdots& \cdots & \cdots & \cdots  & \cdots\cr
  \oint_{\G_{m,N}}\frac{ d\z}{R(\z)} &
\cdots &  \oint_{\G_{m,N}}\frac{\z^{N-1} d\z}{R(\z)} & \overline{ \oint_{\G_{m,N}}\frac{ d\z}{R(\z)}}
& \cdots & \overline{ \oint_{\G_{m,N}}\frac{\z^{N-1} d\z}{R(\z)}}
 &\oint_{\G_{m,N}}\frac{d\z}{(\z-z)R(\z)}\cr
  \oint_{\G_{c,1}}\frac{ d\z}{R(\z)} &
\cdots &  \oint_{\G_{c,1}}\frac{\z^{N-1} d\z}{R(\z)} & \overline{ \oint_{\G_{c,1}}\frac{ d\z}{R(\z)}}
& \cdots & \overline{ \oint_{\G_{c,1}}\frac{\z^{N-1} d\z}{R(\z)}}
 &\oint_{\G_{c,1}}\frac{d\z}{(\z-z)R(\z)}\cr
\cdots &\cdots & \cdots& \cdots & \cdots & \cdots & \cdots \cr
  \oint_{\G_{c,N}}\frac{ d\z}{R(\z)} &
\cdots &  \oint_{\G_{c,N}}\frac{\z^{N-1} d\z}{R(\z)} & \overline{ \oint_{\G_{c,N}}\frac{ d\z}{R(\z)}}
& \cdots & \overline{ \oint_{\G_{c,N}}\frac{\z^{N-1} d\z}{R(\z)}}
&\oint_{\G_{c,N}}\frac{d\z}{(\z-z)R(\z)}\cr
%
%\oint_{\gt_{m,0}}\frac{ d\z}{R(\z)} & \cdots & \oint_{\gt_{m,0}}\frac{\z^{N-1} d\z}{R(\z)} &\overline{ \oint_{\gt_{m,0}}\frac{ d\z}{R(\z)}}
%& \cdots &\overline{ \oint_{\gt_{m,0}}\frac{\z^{N-1} d\z}{R(\z)}}&\Im  \oint_{\gt_{m,0}}\frac{\z^{N} d\z}{R(\z)}&\oint_{\gt_{m,0}}\frac{d\z}{(\z-z)R(\z)}\cr
%
\oint_{\gt}\frac{f(\z)d\z}{R(\z)} & \cdots
&   \oint_{\gt}\frac{\z^{N-1} f(\z)d\z}{R(\z)} &
 \overline{ \oint_{\gt}\frac{f(\z)d\z}{R(\z)}} & \cdots
&  \overline{ \oint_{\gt}\frac{\z^{N-1} f(\z)d\z}{R(\z)}} &\
\oint_{\gt}\frac{f(\z)d\z}{(\z-z)R(\z)}\cr
\end{matrix}\right|,
\ee
where $\G_{m,j}$, $j=1,2,\cdots, N$, and $\G_{c,j}$, $j=1,2,\cdots, N$, denote basic
$\a$ and $\b$ cicles of the corresponding hyperelliptic surface. In the case of our 
countour $\g$, see Fig. \ref{mainandcomp}, 
$\G_{m,j}$, $j=1,2,\cdots, n$, is $\gt_{m,j}^+$,
$\G_{m,j}$, $j=n+1,n+2,\cdots, 2n$, is $\gt_{m,j}^-$,
where $\l^\pm$ denote parts of the contour $\l$ that lie the upper and lower
halfplanes respectively. Similarly, 
$\G_{c,j}$, $j=1,2,\cdots, n$, is $\gt_{c,j}^+$,
$\G_{c,j}$, $j=n+1,n+2,\cdots, 2n$, is $\gt_{c,j}^-$.

According to \cite{TV2}, $D\neq 0$ and 
\eqref{hK}-\eqref{dhdxtdK} are still valid when $D$ and $K$ are given by \eqref{D}, \eqref{K}.
%Hence, so are \eqref{dhdxtdK}. 
Denoting by $K_j$, $j=1,2\cdots,N$, 
the $j$th column in \eqref{D}, and by $Q(z)$ the first
$2n$ entries in the last column of \eqref{K}, we can easily obtain 
\begin{align}
\label{dKdxtgen} 
\frac{\part}{\part x} K(z)=&-\det(K_1,\cdots,K_{N-1},\overline{K_1},\cdots,\overline{K_{N-1}}, 
K_N+\overline{K_N}, Q(z)) \cr
\frac{\part}{\part t} K(z)=&2\det(K_1,\cdots,K_{N-2},K_{N},\overline{K_1},\cdots,\overline{K_{N-2}}, 
K_{N-1}+\overline{K_{N-1}}, \overline{K_{N}},Q(z)) \cr
+&\sum_{j=0}^{4n+1}\a_j\frac{\part}{\part x} K(z)~ .\cr 
\end{align}

\section{Continuity of $h_x$ and $h_t$ across a breaking curve}\label{conthxtbre}

\begin{theorem}\label{equalh}
Let $(x_b,t_b)$ be a regular breaking point and $\a$ be the corresponding breaking point in the 
spectral plane that is  a  double point. Let  $(x_b,t_b)\in l$, where $l$ is
 a breaking curve that separates
regions of genus $2n$ and of genus $2n-2$, $n\in\Z^+$.
If $h^{(2n)}(z;x,t)$  denotes the function $h$  in the genus $2n$ region (on one side of $l$) 
 and 
$h^{(2n-2)}(z;x,t)$ denote the function $h$ in the genus $2n-2$ region (on the other side $l$),
% that has  additional Shwarz symmetrical main arc $\g_{m,n+1}$ in the upper halfplane that emanates from
%the double point $\a$ in the spectral plane when $(x,t)=(x_0,t_0)$. 
then at the point $(x,t)=(x_b,t_b)$ we have
\be\label{hxt=hxt} 
\frac{d}{dx}h^{(2n-2)}(z;x,t)\equiv \frac{d}{dx}h^{(2n)}(z;x,t)~~~{\rm and}~~~\frac{d}{dx}h^{(2n-2)}(z;x,t)\equiv \frac{d}{dx}h^{(2n)}(z;x,t).
\ee

\end{theorem}

\bp
The proof is based on formulae \eqref{dhdxtdK}.
%For technical reasons, it will be conveninet for us to  assume that the pair of main arcs $\g_{m,1}$
%collapses into a pair of double points $\a$  and $\bar\a$. 
We consider the situation when  the pair of main arcs $\g_{m,n}$
collapses into a pair of double points $\a$  and $\bar\a$. That means that
the corresponding branchpoints $\a_{4n-2}$ and $\a_{4n}$    are collapsing into a point $\a$   and the their complex-conjugated branchpoints
$\a_{4n-1}, \a_{4n+1}$ are collapsing into $\bar\a$. It is convenient
to introduce $\d=|\a_{4n-2}(x,t)-\a_{4n}(x,t)|$,
%$\d=\max_{j=1,2}\{|\a_{2j}(x,t)-\a|\}$, 
where $\d\ra 0$.

We first evaluate the $2\times 2$ determinant $D_2$, given by \eqref{DNLS}, with $n=1$
in the limit  $\d\ra 0$. Observe that
%Let $\a=a+ib$ denote the common lomit of $\a_2,\a_4$ as $\d\ra 0$. Then
\be\label{D2}
D_2=\left|\begin{matrix}\oint_{\gt_m}\frac{ d\z}{R(\z)} &\oint_{\gt_m}\frac{\z d\z}{R(\z)} \cr
\oint_{\gt_c}\frac{ d\z}{R(\z)} & \oint_{\gt_c}\frac{\z d\z}{R(\z)}\end{matrix}\right|
=
\left|\begin{matrix}\oint_{\gt_m}\frac{ d\z}{R(\z)} &\oint_{\gt_m}\frac{(\z-\a_2) d\z}{R(\z)} \cr
\oint_{\gt^+_c}\frac{ d\z}{R(\z)} & \oint_{\gt^+_c}\frac{(\z-\a_2) d\z}{R(\z)}\end{matrix}\right|
+
\left|\begin{matrix}\oint_{\gt_m}\frac{ d\z}{R(\z)} &\oint_{\gt_m}\frac{(\z-\bar\a_2) d\z}{R(\z)} \cr
\oint_{\gt^-_c}\frac{ d\z}{R(\z)} & \oint_{\gt^-_c}\frac{(\z-\bar \a_2) d\z}{R(\z)}\end{matrix}\right|=D_2^+ +D_2^-~,
\ee
where  $\nu^\pm$ denote parts of the contour $\nu$ in the upper/lower half-planes respectively. 
Here we  use notation $\g_m$ for $\g_{m,1}$ and $\g_c$ for $\g_{c,1}$.
It is clear that all but $(2,1)$ entries of both determinants $D_2^+ ,D_2^-$ stay bounded  as $\d\ra 0$.
 Using the fact that in the limit $\d\ra 0$
\be\label{RR0}
R(z)=(z-\a)(z-\bar\a)R_0(z)+O(\d)
\ee
provided that $z$ is separated from $\a$ and from $\bar \a$, where 
\be\label{R02}
R_0(z)=\sqrt{(z-\a_0)(z-\bar\a_0)}~,
\ee
we obtain 
\be\label{limgm}
\oint_{\gt_m}\frac{(\z-\a_2) d\z}{R(\z)}=-\frac{2\pi i}{R_0(\bar\a)}(1+o(1))
\ee 
as $\d\ra 0$, where $\a=a+ib$. Using the similar estimate for $\oint_{\gt_m}\frac{(\z-\bar\a_2) d\z}{R(\z)}$,
we finally arrive at
\begin{align}\label{limD2}
D_2= \frac{2\pi i}{2ib|R_0(\a)|^2}&\left[\int_{\g^+_c}\frac{ d\z}{\sqrt{(\z-\a_4)(\z-\a_2)}}-
\int_{\g^-_c}\frac{ d\z}{\sqrt{(\z-\bar\a_4)(\z-\bar\a_2)}}\right] + O(1)\cr
&=\frac{2\pi}{b|R_0(\a)|^2}\ln\d+O(1)\cr
\end{align}
as $\d\ra 0$.

Consider now $D_{2n}$, given by \eqref{DNLS} with $n=2,3,\cdots$,  where the  main arc $\g_{m,n}$ is collapsing into a point
$\a$ when  $(x,t)\ra (x_b,t_b)$.
Rewriting 
\begin{equation}
\label{D2n+2doub}
D_{2n}=(-1)^{n-1}
%D^+_{2n+2}+D^-_{2n+2}=
\left|\begin{matrix} \oint_{\gt_{m,1}}\frac{ d\z}{R(\z)} &\oint_{\gt_{m,1}}\frac{\z d\z}{R(\z)} &
\oint_{\gt_{m,1}}\frac{(\z-\a_*)(\z-\bar\a_*) d\z}{R(\z)} &
\cdots & \oint_{\gt_{m,1}}\frac{\z^{2n-3}(\z-\a_*)(\z-\bar\a_*) d\z}{R(\z)} \cr
\cdots &\cdots & \cdots & \cdots & \cdots \cr
\oint_{\gt_{m,n-1}}\frac{ d\z}{R(\z)} &\oint_{\gt_{m,n-1}}\frac{\z d\z}{R(\z)} &
\oint_{\gt_{m,n-1}}\frac{(\z-\a_*)(\z-\bar\a_*) d\z}{R(\z)} &
\cdots & \oint_{\gt_{m,n-1}}\frac{\z^{2n-3}(\z-\a_*)(\z-\bar\a_*) d\z}{R(\z)} \cr
\oint_{\gt_{c,1}}\frac{ d\z}{R(\z)} &\oint_{\gt_{c,1}}\frac{\z d\z}{R(\z)} &
\oint_{\gt_{c,1}}\frac{(\z-\a_*)(\z-\bar\a_*) d\z}{R(\z)} &
\cdots & \oint_{\gt_{c,1}}\frac{\z^{2n-3}(\z-\a_*)(\z-\bar\a_*) d\z}{R(\z)} \cr
\cdots &\cdots & \cdots & \cdots & \cdots \cr
\oint_{\gt_{c,n-1}}\frac{ d\z}{R(\z)} &\oint_{\gt_{c,n-1}}\frac{\z d\z}{R(\z)} &
\oint_{\gt_{c,n-1}}\frac{(\z-\a_*)(\z-\bar\a_*) d\z}{R(\z)} &
\cdots & \oint_{\gt_{c,n-1}}\frac{\z^{2n-3}(\z-\a_*)(\z-\bar\a_*) d\z}{R(\z)} \cr
\oint_{\gt_{m,n}}\frac{ d\z}{R(\z)} &\oint_{\gt_{m,n}}\frac{\z d\z}{R(\z)} &
\oint_{\gt_{m,n}}\frac{(\z-\a_*)(\z-\bar\a_*) d\z}{R(\z)} &
\cdots & \oint_{\gt_{m,n}}\frac{\z^{2n-3}(\z-\a_*)(\z-\bar\a_*) d\z}{R(\z)} \cr 
\oint_{\gt_{c,n}}\frac{ d\z}{R(\z)} &\oint_{\gt_{c,n}}\frac{\z d\z}{R(\z)} &
\oint_{\gt_{c,n}}\frac{(\z-\a_*)(\z-\bar\a_*) d\z}{R(\z)} &
\cdots & \oint_{\gt_{c,n}}\frac{\z^{2n-3}(\z-\a_*)(\z-\bar\a_*) d\z}{R(\z)} \cr 
\end{matrix}\right|~,
\end{equation}
where $\a_*=\a_{4n-2}$,
and  using \eqref{RR0}, where  
\be\label{R0n}
R_0(z)=\sqrt{\prod_{j=0}^{2n-2}(z-\a_{2j})(z-\overline{\a_{2j}})}~,
\ee
we see that all but the first two entries of the $(2n-1)$th (next to the last) row of $D_{2n}$
are approaching zero as $\d\ra 0$. Taking into account \eqref{limD2} and the fact that all the entries
$(2n,j)$, $j=3,4,\cdots,2n$  of the determinant \eqref{D2n+2doub} are bounded, we obtain
\be\label{limD2n+2}
D_{2n+2}=(-1)^{n-1}D_2\left[ D_{2n-2} +o(1)\right] 
\ee
as $\d\ra 0$, where $ D_{2n-2}$ denotes the determinant built on the main arcs 
$\g_{m,1},\cdots,\g_{m,n-1}$ and the corresponding  complementary
arcs.

Our next step is evaluation of  $\frac{d}{dx}h^{(2n)}(z;x,t)=\frac{R(z)}{D_{2n}}\frac{\part}{\part x} K^{(2n)}(z;x,t)$ in the limit $\d\ra 0$, i.e., when $(x,t)\ra (x_b,t_b)$.
Here $K^{(2n)}(z)=K^{(2n)}(z;x,t)$ denotes $2n+1$ dimensional determinant $K(z)$ given by \eqref{KNLS}.
This evaluation is based on the identity
\be\label{partfrac}
\frac{1}{(\z-z)(\z-\a_*)(\z-\bar\a_*)}=\frac{1}{(z-\a_*)(z-\bar\a_*)}\left[\frac{1}{\z-z}-
\frac{\z+z-2\Re\a_*}{(\z-\a_*)(\z-\bar\a_*)} \right]~, 
\ee
where $\a_*\in\C$ is arbitrary. Using \eqref{partfrac}, the integrand $\frac{1}{(\z-z)R(\z)}$
of the last column of determinant \eqref{KNLS} can be represented as
\be\label{lastcolint}
\frac{1}{(\z-z)R(\z)}=\frac{(\z-\a_*)(\z-\bar\a_*)}{(z-\a_*)(z-\bar\a_*)(\z-z)R(\z)}-
\frac{\z+z-2\Re\a_*}{(z-\a_*)(z-\bar\a_*)R(\z)}~.
\ee
Since the latter term can be eliminated by linear operations  with columns of \eqref{KNLS},
we obtain
$$\frac{\part}{\part x} K^{(2n)}(z)=\frac{(-1)^{n-1}}{(z-\a_*)(z-\bar\a_*)}\times$$
\be\label{Kx2n+2doub}
\hspace{-.8cm}
\left|\begin{matrix} \oint_{\gt_{m,1}}\frac{ d\z}{R(\z)} &\oint_{\gt_{m,1}}\frac{\z d\z}{R(\z)} &
\oint_{\gt_{m,1}}\frac{(\z-\a_*)(\z-\bar\a_*) d\z}{R(\z)} &
\cdots & \oint_{\gt_{m,1}}\frac{\z^{2n-4}(\z-\a_*)(\z-\bar\a_*) d\z}{R(\z)}& 
\oint_{\gt_{m,1}}\frac{ (\z-\a_*)(\z-\bar\a_*)d\z}{(\z-z)R(\z)}
\cr
\cdots &\cdots & \cdots & \cdots & \cdots  & \cdots\cr
\oint_{\gt_{m,n-1}}\frac{ d\z}{R(\z)} &\oint_{\gt_{m,n-1}}\frac{\z d\z}{R(\z)} &
\oint_{\gt_{m,n-1}}\frac{(\z-\a_*)(\z-\bar\a_*) d\z}{R(\z)} &
\cdots & \oint_{\gt_{m,n-1}}\frac{\z^{2n-4}(\z-\a_*)(\z-\bar\a_*) d\z}{R(\z)} &
\oint_{\gt_{m,n-1}}\frac{(\z-\a_*)(\z-\bar\a_*) d\z}{(\z-z)R(\z)}
\cr
\oint_{\gt_{c,1}}\frac{ d\z}{R(\z)} &\oint_{\gt_{c,1}}\frac{\z d\z}{R(\z)} &
\oint_{\gt_{c,1}}\frac{(\z-\a_*)(\z-\bar\a_*) d\z}{R(\z)} &
\cdots & \oint_{\gt_{c,1}}\frac{\z^{2n-4}(\z-\a_*)(\z-\bar\a_*) d\z}{R(\z)} &
\oint_{\gt_{c,1}}\frac{(\z-\a_*)(\z-\bar\a_*) d\z}{(\z-z)R(\z)}
\cr
\cdots &\cdots & \cdots & \cdots & \cdots  & \cdots\cr
\oint_{\gt_{c,n-1}}\frac{ d\z}{R(\z)} &\oint_{\gt_{c,n-1}}\frac{\z d\z}{R(\z)} &
\oint_{\gt_{c,n-1}}\frac{(\z-\a_*)(\z-\bar\a_*) d\z}{R(\z)} &
\cdots & \oint_{\gt_{c,n-1}}\frac{\z^{2n-4}(\z-\a_*)(\z-\bar\a_*) d\z}{R(\z)}
&\oint_{\gt_{c,n-1}}\frac{ (\z-\a_*)(\z-\bar\a_*)d\z}{(\z-z)R(\z)}
 \cr
\oint_{\gt_{m,n}}\frac{ d\z}{R(\z)} &\oint_{\gt_{m,n}}\frac{\z d\z}{R(\z)} &
\oint_{\gt_{m,n}}\frac{(\z-\a_*)(\z-\bar\a_*) d\z}{R(\z)} &
\cdots & \oint_{\gt_{m,n}}\frac{\z^{2n-4}(\z-\a_*)(\z-\bar\a_*) d\z}{R(\z)} &
\oint_{\gt_{m,n}}\frac{(\z-\a_*)(\z-\bar\a_*) d\z}{(\z-z)R(\z)}
\cr 
\oint_{\gt_{c,n}}\frac{ d\z}{R(\z)} &\oint_{\gt_{c,n}}\frac{\z d\z}{R(\z)} &
\oint_{\gt_{c,n}}\frac{(\z-\a_*)(\z-\bar\a_*) d\z}{R(\z)} &
\cdots & \oint_{\gt_{c,n}}\frac{\z^{2n-4}(\z-\a_*)(\z-\bar\a_*) d\z}{R(\z)} 
&\oint_{\gt_{c,n}}\frac{ (\z-\a_*)(\z-\bar\a_*)d\z}{(\z-z)R(\z)}
\cr 
\end{matrix}\right|.
\ee
Let $M^{(2n-2)}(z)$ denote the minor of \eqref{Kx2n+2doub} that consists of the first $2n-2$
rows and the last $2n-2$ columns. 
Choosing $\a_*=\a_{4n-2}$, we can replace the factor 
$\frac{ (\z-\a_*)(\z-\bar\a_*)}{R(\z)}$ in all the integrands of the minor $M^{(2n-2)}(z)$
by $\frac{ 1}{R_0(\z)}$ with the accuracy $O(\d)$ as $\d\ra 0$. So,
$M^{(2n-2)}(z)=\frac{\part}{\part x} K^{(2n-2)}(z)+O(\d)$.
Note also that, for any fixed $z\ne \a$,  all but the first two enties of the $(2n-1)$st row of \eqref{Kx2n+2doub}
have the order  $O(\d)$, and all  but the first two enties of the last row of \eqref{Kx2n+2doub}
are bounded as $\d\ra 0$ . Thus, applying to \eqref{Kx2n+2doub} the arguments of  \eqref{D2n+2doub}, we obtain
\be\label{ddxK2n2n-2}
\frac{\part}{\part x} K^{(2n)}(z)=\frac{(-1)^{n-1}D_2\frac{\part}{\part x}K^{(2n-2)}(z)}{(z-\a_{4n-2})(z-{\bar\a_{4n-2}})}+O(\d)
\ee
as $\d\ra 0$, which holds uniformly in $z$ on compact subsets of $\C\setminus\{\a,\bar\a\}$.
Now, according to \eqref{dKdxn}, \eqref{limD2n+2}, \eqref{R0n} and \eqref{ddxK2n2n-2},  we have
\begin{align}\label{hx=hx}
\left.\frac{d}{dx}h^{(2n)}(z;x,t)\right|_{(x,t)=(x_b,t_b)}&=\lim_{\d\ra 0}
\frac{(-1)^{n-1}D_2R(z)\frac{\part}{\part x}K^{(2n-2)}(z;x,t)}{(z-\a_{4n-2})(z-{\bar\a_{4n-2}})D_{2n}}\cr
&=\left. \frac{R_0(z)}{D_{2n-2}}\frac{\part}{\part x} K^{(2n-2)}(z;x,t)
=\frac{d}{dx}h^{(2n-2)}(z;x,t)\right|_{(x,t)=(x_b,t_b)}
\end{align}
for any $z\in \C$. 
%where $K^{(2n-2)}(z;x,t)$ denotes the determinant \eqref{KNLS} built on the main arcs 
%$\g_{m,2},\cdots,\g_{m,n}$ and the corresponding  complex-conjugated arcs in the lower halfplane. 
Thus, the first equation in \eqref{hxt=hxt} is proven.

We now turn to the second equation in \eqref{hxt=hxt}. Similarly to 
\eqref{Kx2n+2doub}, we represent  $\frac{\part}{\part t} K^{(2n)}(z)$ as
$$\frac{\part}{\part t} K^{(2n)}(z)=\frac{(-1)^{n-1}}{(z-\a_*)(z-\bar\a_*)}
\times$$
\be\label{dKdtndoub}
\hspace{-.6cm}
\left| \begin{matrix}
\oint_{\gt_{m,1}}\frac{ d\z}{R(\z)} &\oint_{\gt_{m,1}}\frac{\z d\z}{R(\z)} &\cdots &
\oint_{\gt_{m,1}}\frac{\z^j d\z}{R_0(\z)} &
\cdots & \oint_{\gt_{m,1}}\frac{[\z-\hf\sum_{j=0}^{4n+1}\a_j]\z^{2n-2} d\z}{R(\z)}& 
\oint_{\gt_{m,1}}\frac{(\z-\a_*)(\z-\bar\a_*) d\z}{(\z-z)R(\z)}
\cr
\cdots &\cdots & \cdots & \cdots & \cdots  & \cdots &\cdots\cr
 \oint_{\gt_{m,n-1}}\frac{ d\z}{R(\z)} &\oint_{\gt_{m,n-1}}\frac{\z d\z}{R(\z)} &\cdots &
\oint_{\gt_{m,n-1}}\frac{\z^j d\z}{R_0(\z)} &
\cdots & \oint_{\gt_{m,2}}\frac{[\z-\hf\sum_{j=0}^{4n+1}\a_j]\z^{2n-2} d\z}{R(\z)} &
\oint_{\gt_{m,n-1}}\frac{(\z-\a_*)(\z-\bar\a_*) d\z}{(\z-z)R(\z)}
\cr
\oint_{\gt_{c,1}}\frac{ d\z}{R(\z)} &\oint_{\gt_{c,1}}\frac{\z d\z}{R(\z)} &\cdots &
\oint_{\gt_{c,1}}\frac{z^jd\z}{R_0(\z)} &
\cdots & \oint_{\gt_{c,1}}\frac{\z^{2n-2} d\z}{R(\z)} &
\oint_{\gt_{c,1}}\frac{(\z-\a_*)(\z-\bar\a_*) d\z}{(\z-z)R(\z)}
\cr
\cdots &\cdots & \cdots & \cdots & \cdots  &\cdots & \cdots\cr
\oint_{\gt_{c,n-1}}\frac{ d\z}{R(\z)} &\oint_{\gt_{c,2}}\frac{\z d\z}{R(\z)} &\cdots &
\oint_{\gt_{c,n-1}}\frac{\z^j d\z}{R_0(\z)} &
\cdots & \oint_{\gt_{c,n-1}}\frac{[\z-\hf\sum_{j=0}^{4n+1}\a_j]\z^{2n-2} d\z}{R(\z)}
&\oint_{\gt_{c,n-1}}\frac{(\z-\a_*)(\z-\bar\a_*) d\z}{(\z-z)R(\z)}
 \cr
 \oint_{\gt_{m,n}}\frac{ d\z}{R(\z)} &\oint_{\gt_{m,n}}\frac{\z d\z}{R(\z)} &\cdots &
\oint_{\gt_{m,n}}\frac{ \z^jd\z}{R_0(\z)} &
\cdots & \oint_{\gt_{m,n}}\frac{[\z-\hf\sum_{j=0}^{4n+1}\a_j]\z^{2n-2} d\z}{R(\z)} &
\oint_{\gt_{m,n}}\frac{(\z-\a_*)(\z-\bar\a_*) d\z}{(\z-z)R(\z)}
\cr 
\oint_{\gt_{c,n}}\frac{ d\z}{R(\z)} &\oint_{\gt_{c,n}}\frac{\z d\z}{R(\z)} &\cdots &
\oint_{\gt_{c,n}}\frac{\z^j d\z}{R_0(\z)} &
\cdots & \oint_{\gt_{c,n}}\frac{[\z-\hf\sum_{j=0}^{4n+1}\a_j]\z^{2n-2} d\z}{R(\z)} 
&\oint_{\gt_{c,n}}\frac{(\z-\a_*)(\z-\bar\a_*) d\z}{(\z-z)R(\z)}
\cr 
\end{matrix}\right|,
\ee
where $j=0,1,\cdots,2n-4$ and $\a_*\in\C$ is arbitrary.
Using the  identity
\be\label{polyident}
\left( \z-\hf\sum_{j=0}^{4n-3}\a_j \right)(\z-\a^*)(\z-{\bar \a^*})\z^{2n-4}= 
\z^{2n-1}-\frac{\z^{2n-2}}{2}\sum_{j=0}^{4n+1}\a_j +O(\z^{2n-3})~,
\ee
where $\a^*=\hf(\a_{4n-2}+\a_{4n})$, 
%and $\bar\a=\a_{4n-1}=\a_{4n+1}$, 
we can reduce the integrand in the $(2n-1)$st
(next to the last) column of the
latter determinant to 
\be\label{colKt}
\frac{\left[\z-\hf\sum_{j=0}^{4n-3}\a_j)\right](\z-\a^*)(\z-{\bar \a^*})\z^{2n-4} }{R(\z)}=
\frac{\left[\z-\hf\sum_{j=0}^{4n-3}\a_j)\right]\z^{2n-4} }{R_0(\z)}+O(\d) 
\ee
as $\d\ra 0$. The latter estimate is valid if $\z\ne\a$, $\z\ne\bar\a$  uniformly on 
compact subsets of
$\C\setminus \{\a,\bar\a\}$. Thus the integrand in all but the last integral in the $(2n-1)$st
 column can be replaced by $\frac{\left[\z-\hf\sum_{j=0}^{4n-3}\a_j)\right]\z^{2n-4} }{R_0(\z)}$
with accuracy $O(\d)$. We also note that the last integral in this column is bounded.
Denoting the latter determinant by $\hat K$ and applying to it the same arguments as we applied to \eqref{Kx2n+2doub}, and also using \eqref{colKt}, we obtain
\be\label{hatK}
\hat K=\hf D_2\frac{\part}{\part t} K^{(2n-2)}(z;x_b,t_b)~.
%\left[ \frac{\part}{\part t} K^{(2n-2)}(z;x_0,t_0)+o(1)\right]~. 
\ee
Then \eqref{dKdtn}, \eqref{D2n+2doub} and \eqref{hatK} yield
\be\label{ht=ht}
\frac{d}{dt}h^{(2n)}(z;x,t)\left|_{(x,t)=(x_b,t_b)}=\frac{R_0(z)}{D_{2n-2}} \frac{\part}{\part t} K^{(2n-2)}(z;x,t)\right|_{(x,t)=(x_b,t_b)}=\frac{d}{dt}h^{(2n-2)}(z;x,t)\left|_{(x,t)=(x_b,t_b)}.\right.
\ee

In the remaining case $n=1$, expressions \eqref{Kx2n+2doub} and \eqref{dKdtndoub} become
\begin{align}\label{hxt2=0}
h^{(2)}_x(z;x_b,t_b)&=-\lim_{\d\ra 0}\frac{R_0(z)}{D_2}\left[ D_2+
\left|\begin{matrix}\oint_{\gt_m}\frac{ d\z}{(\z-z)R_0(\z)} &\oint_{\gt_m}\frac{ d\z}{R(\z)} \cr
\oint_{\gt_c}\frac{ d\z}{(\z-z)R_0(\z)} & \oint_{\gt_c}\frac{ d\z}{R(\z)}\end{matrix}\right|\right] 
= -R_0(z)\cr
h^{(2)}_t(z;x_b,t_b)&=2\lim_{\d\ra 0}\frac{R_0(z)}{D_2}\left[
\left|\begin{matrix}\oint_{\gt_m}\frac{ d\z}{(\z-z)R_0(\z)} &\oint_{\gt_m}\frac{\z d\z}{R(\z)} \cr
\oint_{\gt_c}\frac{ d\z}{(\z-z)R_0(\z)} & \oint_{\gt_c}\frac{\z d\z}{R(\z)}\end{matrix}\right|
-(z-2a_0)D_2\right]=-2(z-2a_0)R_0(z)\cr~,
\end{align}
where $a_0=\Re\a_0$.
According to Corollary  4.4 from \cite{TVZ1}, in the genus zero region 
\be\label{hxt0}
h^{(0)}_x(z)=-R_0(z)~~~~{\rm and}~~~~h^{(0)}_t(z)=-2(z+a_0)R_0(z)~.
\ee
These expressions, combined with \eqref{hxt2=0}, complete the proof of the theorem
for $n=1$.
\ep

\section{Regular continuation principle}\label{regcontpr}

To prove the regular continuation principle, we need Theorem \ref{equalh} and certain facts about the geometry of breaking curves. Namely, we need to prove that any regular nondegenerate breaking point lies 
on a smooth breaking curve and that any  regular degenerate breaking point is an isolated point in the $(x,t)$ plane.

\begin{theorem}\label{brcurve}
If $(x_b,t_b)$ is a regular nondegenerate  breaking point, then there exists a breaking curve $l$
passing through
$(x_b,t_b)$. Moreover, $l$ is smooth and defined uniquely.
\end{theorem}

\bp
If $(x_b,t_b)$ is a regular nondegenerate breaking point, 
then $\exists z_0\in\g$, such that $h'(z_0)$ but 
$h''(z_0)\neq 0$. Thus, $z_0$ and $(x_b,t_b)$ satisfy the system
\be
\begin{cases}\label{dpteq}
h'(z;x,t)&=0  \cr
\Im h(z;x,t)&=0\cr
\end{cases}
\ee
of three real equation for four real variables $u,v x,t$, where $z=u+iv$.

According to Theorem \ref{Imhthxneq0} below, if $z\not\in \R$ and if $z$ is not a branchpoint,
then $\Im h_x(z)$ and $\Im h_t(z)$ cannot be zero simultaneously. Let us assume,
for example, that $\Im h_t(z_0)\neq 0$. Then, using the Cauchy-Riemann equations and the fact 
that $h'(z_0)=0$, the Jacobian of the system \eqref{dpteq} at $(z_0,x_b,t_b)$ is
%we obtain that at the point $z=z_0$
\be\label{jac1}
%\hspace{-.6cm}
\left| \begin{matrix}
\frac{\part}{\part u} \Re h'&\frac{\part}{\part v} \Re h'&\frac{\part}{\part t} \Re h'\cr
\frac{\part}{\part u} \Im h'&\frac{\part}{\part v} \Im h'&\frac{\part}{\part t} \Im h'\cr
\frac{\part}{\part u} \Im h&\frac{\part}{\part v} \Im h&\frac{\part}{\part t} \Im h\cr
\end{matrix}
\right|=
\left|h''(z_0)\right|^2\cdot \Im h_t(z_0)\neq 0~.
\ee
Now, the Implicit Function Theorem completes the proof.
\ep

\begin{corollary}\label{corbcurve}
Let $(x_b,t_b)$ be a regular nondegenerate  breaking point and $z_0$ be the corresponding (double)
breaking point in the spectral plane. Then there exists a unique smooth curve $\l$, so that $z_0$
varies along $\l$ as the correspondin breaking point $(x_b,t_b)$ varies along $l$.
\end{corollary}

To prove Theorem \ref{hthxneq0}, we first need the following lemma.

\begin{lemma}\label{abmap}
If $\Rscr$ is an hyperelliptic Riemann surface of  genus $g>0$ and if $P_0,P_1$ are two
fixed points on $\Rscr$, then there exists a holomorphic differential $\o$
on $\Rscr$ such that $\int_{P_0}^{P_1}\o\neq 0$. Here we assume that the integral is single-valued,
i.e., the contour of integration does not cross any $\a$ or $\b$ cycle of $\Rscr$.
\end{lemma}

\bp 
Suppose the converse is true. Then for ${P_0}$ and ${P_1}$ the Abel map is trivial.
By Abel's Theorem,  $P_1-P_0$ is a principle divisor, i.e.,  there exists a meromorphic function
$\phi$ on $\Rscr$ with the only pole at $P_0$ and the only zero at $P_1$, both the pole and the
zero are simple. Then $\phi$ provides a diffeomorphism between $\Rscr$ and the Riemann sphere,
which is a contradiction to the fact that $g>0$.  
\ep

\begin{theorem}\label{hthxneq0}
Let $h(z)$ be defined by \eqref{hform} with some $N=2n$, $n\in\N$. If $z$ is not a branchpoint 
$\a_j$, $j=0,1,\cdots,4n+1$, then
\be\label{|ht|+|hx|}
| h_x(z)|+| h_t(z)|\neq 0.
\ee
\end{theorem}
\bp
Let us fix some $z$.
In the case $n=0$, \eqref{|ht|+|hx|} follows from \eqref{hxt0}. In the case $n>0$, according
to \eqref{dhdxtdK}, $| h_x(z)|+|h_t(z)|=0$ is equivalent to 
\be\label{dKdxt=0}
\left|\frac{\part}{\part x} K(z)\right|+\left|\frac{\part}{\part t} K(z)\right|=0~.
\ee
Let us assume that \eqref{dKdxt=0} is true.  
Consider $\frac{\part}{\part x} K(z),~\frac{\part}{\part t} K(z)$ given by
\eqref{dKdxtgen}. If the period vector $Q(z)=$
\be\label{Q(z)}
\Col\left( \oint_{\G_{m,1}}\frac{d\z}{(\z-z)R(\z)},\cdots,\oint_{\G_{m,n}}\frac{d\z}{(\z-z)R(\z)},
\oint_{\G_{c,1}}\frac{d\z}{(\z-z)R(\z)},\cdots,\oint_{\G_{m,n}}\frac{d\z}{(\z-z)R(\z)}
\right) 
\ee
of the meromorphic differential $\eta=\frac{ d\z}{(\z-z)R(\z)}$ on 
the Riemann surface $\Rscr$ is different from zero,
%$Q(z)\neq 0$, 
then $Q(z)$ is a nontrivial linear combination 
of columns $K_j$ and their complex conjugates from the determinant  $\frac{\part}{\part x} K(z)$.
Substituting this linear combination into $\frac{\part}{\part t} K(z)$, we see that, according to
\eqref{dKdxt=0}, a nontrivial linear combination of columns of  determinant $D$,
given by \eqref{D}, is zero. Since $D\neq 0$, the obtained contradiction shows that 
\eqref{|ht|+|hx|} is true. To complete the proof, it remains to show that $Q(z)\neq 0$.

Note that $\eta$ is an
%$Q(z)$ is the $2N$ dimensional vector of $\a$ and $\b$ periods of the 
abelian differential of the third kind $\eta=\frac{ d\z}{(\z-z)R(\z)}$ 
(a meromorphic differential with
nonzero residues) on  $\Rscr$. Riemann bilinear relation for $\eta$ is
(see, for example, \cite{Belo}),
\be\label{bilin3}
\sum_{k=1}^N(A'_kB_K-A_kB'_k)=2\pi i \sum c_j\int_{P_0}^{P_j}\o~,
\ee
where: $\o$ is an arbitrary holomorphic differential on $\Rscr$ with $\a$ and $\b$ periods
$\{A'_k,B'_k\}$ respectively; $\{A_k,B_k\}$ are $\a$ and $\b$ periods of $\eta$ respectively;
$P_0$ is an arbitrary point on $\Rscr$; $P_j$ are  the poles of $\eta$ in $\Rscr$ and 
$c_j$ are their residues; the summation in the right hand side of \eqref{bilin3} is taken
over all the poles; a single-valued branch of the (multi-valued) integral 
$\int_{P_0}^{P_j}\o$ is taken in the right hand side of \eqref{bilin3}, i.e.,
integration contours  do not cross any main or any complementary arc except of $\g_{m,0}$
(that has endpoints $\a_0$ and $\bar\a_0$). 
Since $z$ is not a branchpoint, $\eta$ has two simple poles at $P_1=z$ on the main sheet and $P_2=z$ on the secondary sheet of $\Rscr$ with the residues $c_1=\frac{1}{R(z)}$ and $c_2=-\frac{1}{R(z)}$.
Choosing $P_0=\a_0$ and using the fact that all the   $\a$ and $\b$ periods of $\eta $ are zero,
we can rewrite \eqref{bilin3} as 
\be\label{bilin3a}
\frac{1}{R(z)}\left[\int_{P_0}^{P_1}\o + \int_{P_2}^{P_0}\o\right] =0
\ee
Since on the secondary sheet $\o(\z)=-\o(\tilde \z)$, where $\tilde \z$ is the projection of
$\z$ on the main sheet, equation \eqref{bilin3a} becomes
\be\label{bilin3b}
\int_{\a_0}^{z}\o =0~,
\ee
where the contour of integration lies on the main sheet. Note that \eqref{bilin3b} holds
for all the basic holomorphic differentials of $\Rscr$. However, this is contradicts 
Lemma \ref{abmap}. The proof is completed.
\ep

The proof of Theorem \ref{hthxneq0} can be slightly adjusted for the following statement.

\begin{theorem}\label{Imhthxneq0}
Let $h(z)$ be defined by \eqref{hform} with some $N=2n$, $n\in\N$. If $z\not\in\R$ and $z$ 
 is not a branchpoint 
$\a_j$, $j=0,1,\cdots,4n+1$, then
\be\label{Im|ht|+|hx|}
|\Im h_x(z)|+|\Im h_t(z)|\neq 0.
\ee
\end{theorem}

\bp 
Let us fix some $z$.
In the case $n=0$ \eqref{Im|ht|+|hx|} follows from \eqref{hxt0}. 
Consider the case $n>0$.
Since $h_x,h_t$ are Schwarz symmetrical, we have
\be \label{Imhxt}
\Im h_x(z)=-\hf i \left[ h_x(z)-h_x(\bar z)\right],~~~
\Im h_t(z)=-\hf i \left[ h_t(z)-h_t(\bar z)\right]~.
\ee
Then $\Im h_x,t(z)$ are given by 
\eqref{dhdxtdK} and  \eqref{dKdxtgen}, where the last column $Q(z)$ of the periods of the
meromorphic differential $\eta=\frac{ d\z}{(\z-z)R(\z)}$ in \eqref{dKdxtgen} is replaced 
by the column $\tilde Q(z,\bar z)$ of the periods of the
meromorphic differential 
\be\label{teta}
-\frac{i}{2}\teta=-\frac{i}{2}\left[ \frac{R(z) d\z}{(\z-z)R(\z)}- \frac{R(\bar z)d\z}{(\z-\bar z)R(\z)}\right] ~.
\ee
Following the arguments of Theorem  \ref{hthxneq0}, it is sufficeint to prove that 
the vector $\tilde Q(z,\bar z)$ is not equal to zero for any $z\not\in\R$, which 
 is also not a branchpoint. 

Assume that for some $z$, satisfying the requirements of the theorem, $\tilde Q(z,\bar z)=0$. 
Since $\teta$ is an abelian differential of the third kind on $\Rscr$, 
the right hand side of \eqref{bilin3} is zero for any holomorphic differential $\o$.
The differential $\teta$ has simple poles at $\z=z$ and $\z=\bar z$ with residues $\pm 1$
respectively. So, the contribution of these two poles to the right hand side of 
\eqref{bilin3} is $2\pi i  \int_{\bar z}^z \o$. The remaining two poles  $\z=z$ and $\z=\bar z$
on the second sheet of $\Rscr$ give exactly the same contribution. Thus, the Riemann
bilinear relation implies 
\be\label{bilin3c}
\int_{\bar z}^z \o=0
\ee
for all the holomorphic differentials on $\Rscr$. The obtained contradiction with 
Lemma \ref{abmap} completes the proof. 
\ep

Let $z_0$ be the breaking point on the spectral plane that corresponds to a 
regular breaking point $(x_b,t_b)$. If $(x_b,t_b)$ is a degenerate breaking point
then, according to \eqref{degdeg}, the degree of $z_0$ is greater than two, so that
$h''(z_0)=0$. 

\begin{theorem}\label{degisol}
A  regular degenerate breaking point $(x_b,t_b)$ is an isolated point
in the $x,t$-plane, that is, there exists a neighborhood of $(x_b,t_b)$ that is free
of  other degenerated breaking points.
\end{theorem}

\bp A) Let us first consider the case when $z_0$ is not a branchpoint. Then there
exists some  $m=3,4,\cdots$, such that
%Let us assume that at the breaking point $(x_b,t_b)$ we have 
$h^{(k)}(z_0)=0$,
$k=1,2,\cdots, m-1$, but $h^{(m)}(z_0)\neq 0 $, so that $z_0$ and
$(x_b,t_b)$ satisfy the system of $2m-1$ real equations
\be
\begin{cases}\label{hipteq}
h^{(k)}(z;x,t)&=0,~~~~ k=1,2,\cdots, m-1 \cr
\Im h(z;x,t)&=0\cr
\end{cases}
\ee
for four real variables $u,v, x,t$, where $z=u+iv$. Consider the subsystem
%Let us assume that the system  \eqref{hipteq} holds on some curve $l$, $z_0\in l$.
%, that is a part of the boundary of $G_0$. That means, in particular, that equations  
\be
\begin{cases}\label{hipteqmod}
h^{(m-1)}(z;x,t)&=0, \cr
h'(z;x,t)&=0~,\cr
\Im h(z;x,t)&=0~\cr
\end{cases}
\ee
%hold on $l$. 
of \eqref{hipteq}, which has a Jacoby matrix
\be\label{jac2}
%\hspace{-.6cm}
\left( \begin{matrix}
\frac{\part}{\part u} \Re h^{(m-1)}&\frac{\part}{\part v} \Re h^{(m-1)}&\frac{\part}{\part x} \Re h^{(m-1)}&\frac{\part}{\part t} \Re h^{(m-1)}\cr
\frac{\part}{\part u} \Im h^{(m-1)}&\frac{\part}{\part v} \Im h^{(m-1)}&\frac{\part}{\part x} \Im h^{(m-1)}&\frac{\part}{\part t} \Im h^{(m-1)}\cr
\frac{\part}{\part u} \Re h'&\frac{\part}{\part v} \Re h'&\frac{\part}{\part x} \Re h'&\frac{\part}{\part t} \Re h'\cr
\frac{\part}{\part u} \Im h'&\frac{\part}{\part v} \Im h'&\frac{\part}{\part x} \Im h'&\frac{\part}{\part t} \Im h'\cr
\frac{\part}{\part u} \Im h&\frac{\part}{\part v} \Im h&\frac{\part}{\part t} \Im h&\frac{\part}{\part t} \Im h\cr
\end{matrix}
\right)~.
\ee
is the Jacoby matrix  of system \eqref{hipteqmod}. 
At the point $z=z_0$, similarly to \eqref{jac1},  the $2\times 2$ minor in the
upper left corner of \eqref{jac2}  is equal to  $\left|h^{(m)}(z_0)\right|^2\neq 0$, whereas 
the $3\times 2$ block in the lower left corner is a zero matrix.  Theorem \ref{Imhthxneq0} implies 
that the $3\times 2$ block
\be\label{rlblock}
%\hspace{-.6cm}
\left( \begin{matrix}
\frac{\part}{\part x} \Re h'&\frac{\part}{\part t} \Re h'\cr
\frac{\part}{\part x} \Im h'&\frac{\part}{\part t} \Im h'\cr
\frac{\part}{\part t} \Im h&\frac{\part}{\part t} \Im h\cr
\end{matrix}
\right)
\ee
is of at least rank $\r=1$. 
According to the Implicit Function Theorem,  it is sufficient to show that 
the latter block has rank $\r=2$ in order to prove  the theorem. To complete the proof, we assume $\r=1$ and obtain a contradiction.

Let us first obtain a contradiction in the case when 
$h(z)$ at $z=z_0$ is given by \eqref{hform} with 
$n=0$.
% i.e., when the curve $l$ (and $z_0$) is on the boundary of a genus zero region $G_0$. 
In this case  $h_x(z)$ and $h_t(z)$ are given by \eqref{hxt0}, so that 
\be\label{hxztz}
h_{xz}=-\frac{z-a}{R(z)},~~~~h_{tz}=-2\frac{z^2-a^2}{R(z)}-2R(z),
\ee
where $R(z)=\sqrt{(z-\a)(z-\bar \a)}$ and $\a=a+ib$. Since for arbitrary $f$ and $g$
\be\label{Imbarfg}
\left| \begin{matrix}
\Re f & \Re g\cr
 \Im f& \Im g\cr
\end{matrix}
\right|=\Im (\bar f g)~,
\ee
the assumption $\r=1$ implies $\Im \bar h_{xz}h_{tz}=0$. Direct calculation yields
\be\label{Imbarhzxhzt}
\Im h_{xz}(z)h_{tz}(z)=\frac{2\Im z \left[ 2|z-a|^2-b^2\right] }{|z-\a|^2}~.
\ee
Since $\Im z_0>0$, the point $z_0$ must be on the upper semicircle 
\be\label{b}
|z-a|=\frac{b}{\sqrt{2}}.
\ee

Now, let us show that 
\be\label{detg0}
\left| \begin{matrix}
\Im h_x(z) & \Im h_t(z)\cr
 h_{xz}(z)&  h_{tz}(z)\cr
\end{matrix}
\right|=\frac{2}{R(z)}
\left| \begin{matrix}
\Im R(z) & (u+a)\Im R(z)+v\Re R(z)\cr
 z-a&  (z-a)^2+b^2+z^2-a^2\cr
\end{matrix}
\right|\neq 0
\ee
for any $z=u+iv$ with $u>0$ satisfying \eqref{b}. Substituting \eqref{b} into the determinant
in the right hand side of  \eqref{detg0} yield
\be\label{detg0a}
(z-a)\left[3(u-a)\Im R(z)-v\Re R(z) \right]=(z-a)\Im \left[[3(u-a)-iv]  R(z)\right]~.
\ee
Moreover, \eqref{b} yields
\be\label{detg0b}
R(z)=\sqrt{(z-a)^2+b^2}=\sqrt{z-a}\sqrt{3(u-a)-iv}~,
\ee
which, together with \eqref{detg0a}, yield 
\be\label{detg0c}
\left| \begin{matrix}
\Im h_x(z) & \Im h_t(z)\cr
 h_{xz}(z)&  h_{tz}(z)\cr
\end{matrix}
\right|=\frac{2(z-a)}{R(z)}\Im\left[ \left( 3(u-a)-iv\right)^{\frac{3}{2}}
\left( (u-a)-iv\right)^{\frac{1}{2}}\right] 
\ee
To prove $\r=2$, we need to prove 
\be\label{arg1}
3\arg \left( 3(u-a)-iv\right)+\arg \left( (u-a)-iv\right) \neq 2\pi m
\ee
for any $m\in\Z$, where $\th=\arg \left( (u-a)-iv\right)$ varies between $0$ and $\pi$. Equation \eqref{arg1} can be rewrited as
\be\label{arg2}
\phi(\th)=\th+3\tan^{-1}\left(- \frac{\tan \th}{3}\right) \neq 2\pi m~
\ee
if $\th\le \pt$; if $\th>\pt$, we need to subtract $3\pi$ from this expression.
Notice that $\phi(0)=0$, $\phi(\pt)=-\pi$ and $\phi(\pt)=-2\pi$
and  $\phi(\th)$ is monotonically decreasing since
\be\label{phi'}
\phi'(\th)=-8\frac{\sin^2\th}{8\cos^2\th+1}<0~.
\ee
So, inequality \eqref{arg2} holds for all $\th\in(0,\pi)$. In the case $n=0$,
the proof is completed.

In the case of a positive genus $N=2n$, derivatives $h_x$ and $h_t$ are 
given by  \eqref{dhdxtdK}, where $D$
$\frac{\part}{\part x} K(z)$ and $\frac{\part}{\part t} K(z)$ are given by 
\eqref{D} and  \eqref{dKdxtgen} respectively. Then 
\be\label{hzxtn}
h_{xz}(z)=\frac{1}{D} \hat K_x(z),~~~
h_{tz}(z)=\frac{1}{D} \hat K_t(z),
\ee
where $\hat K_x(z),\hat K_t(z)$ are obtained from determinants \eqref{dKdxtgen} 
respectively by replacing the last column $Q(z)$ with the column
$\frac{d}{dz}\left( R(z)Q(z)\right)$ (in $\hat K_x(z),\hat K_t(z)$ the subscript does not
mean  differentiation).

Let us first prove that   $h_{xz}(z), h_{tz}(z)$ cannot be zero simultaneously
for any $z\in\Rscr$. If vector $\frac{d}{dz}\left( R(z)Q(z)\right)\neq 0$, the proof 
is the same as for $h_x,h_t$ in Theorem \ref{hthxneq0}. In the case vector
$\frac{d}{dz}\left( R(z)Q(z)\right)=0$,  we consider  differential 
$\eta=\frac{d}{dz}\left( \frac{R(z)}{\z-z}\right) \frac{d\z}{R(\z)}$ on $\Rscr$. Second order poles
at $\z=z$ on the main and secondary sheets of $\Rscr$ are the only poles of $\eta$. It is
an abelian differential of the second kind  since its residues are zeroes. 
Riemann bilinear relation for $\eta$ and an arbitrary meromorphic differential $\o$ on $\Rscr$ is
given by (see, for example, \cite{FK})
\be\label{bilin2}
\sum_{k=1}^N(A'_kB_K-A_kB'_k)=2\pi i \sum_{P} \Res u\o~,
\ee
where the summation is taken over all the poles $P$ of the meromorphic function 
$u=\int \eta$ and of the meromorphic differential $\o$. Here $\{A'_k,B'_k\}$ are
 $\a$ and $\b$ periods of $\o$
 respectively and $\{A_k,B_k\}$ are $\a$ and $\b$ periods of $\eta$ respectively.
Take $\o$ to be a holomorphic differential.
Since all the periods of $\eta$ are zero and residues of $u\o$ at $z$ are the same on the
both sheets of $\Rscr$,
we can reduce \eqref{bilin2} to 
\be\label{bilin2a}
\left. \Res (u\o)\right|_{\z=z}=0~,
\ee
where $z$ is on the main sheet. Since $\Res u|_{\z=z}=1$ and $\o$ is any holomorhic differential,
we obtain a contradiction. Thus, the second row in the determinant 
\be\label{detgnn}
\left| \begin{matrix}
\Im h_x(z) & \Im h_t(z)\cr
 h_{xz}(z)&  h_{tz}(z)\cr
\end{matrix}
\right|
\ee
is not zero.

Since both rows of the determinant \eqref{detgnn} are nonzero (for every $z\in\Rscr$ that is not a 
branchpoint), it is sufficient to show that for any $\x\in\C$ and any $z\in\Rscr$, the vector 
\be\label{Vxz}
V(\x,z)= (V_1(\x,z),V_2(\x,z))=\left( \Im h_x(z)-\x h_{xz}(z), \Im h_t(z)-\x h_{tz}(z)\right)
\ee
is not zero. Components of $V(\x,z)$ can be represented as
\be\label{v12}
V_1(\x,z)=\frac{1}{D} \hat K_1(\x,z),~~~~V_2(\x,z)=\frac{1}{D} \hat K_2(\x,z),
\ee
where determinants $\hat K_{1,2}(\x,z)$ are obtained from determinants
$\frac{\part}{\part x} K(z)$ and $\frac{\part}{\part t} K(z)$ in
\eqref{dKdxtgen} respectively by replacing
the last column $Q(z)$ with 
\be\label{Zxz}
Z(\x,z)=-\frac{i}{2}[R(z)Q(z)-R(\bar z)Q(\bar z)]-\x\frac{d}{dz}\left( R(z)Q(z)\right)~.
\ee
If the  vector $Z(\x,z)\neq 0$ then, as in the proof  of Theorem \ref{hthxneq0}, 
we can establish that $V_1(\x,z)$ and $V_2(\x,z)$ cannot be zero simultaneously.
In the remaining case $Z(\x,z)=0$ we consider the meromorphic differential
\be\label{etadiff}
\eta= \left\lbrace  -\frac{i}{2}\left( \frac{R(z)}{\z-z}-\frac{R(\bar z)}{\z-\bar z}\right) -\x\frac{d}{dz}\left(\frac{R(z)}{\z-z} \right)\right\rbrace  \frac{d\z}{R(\z)}
\ee
on $\Rscr$. This is an abelian differential of the third kind with poles at $\z=z$
and $\z=\bar z$ on the both sheets of $\Rscr$. The residues of $\eta$ at  $\z=z$
and $\z=\bar z$ (on the main sheet) are $-\frac{i}{2}$ and $\frac{i}{2}$ respectively.
Thus, we can repeat the arguments of Theorem \ref{hthxneq0} to prove that  
 $Z(\x,z)=0$ is not possible. 

B) Let us now consider the case when  $z_0$ is a branchpoint. If $h(z)$ 
at $z=z_0$ is given by \eqref{hform} with 
$n=0$ (genus zero case), the statement of the theorem was proven in 
\cite{TVZ3}, Lemma 3.21. Otherwise, we assume $n>0$. Note that if $z_0$ is a
branchpoint, say, $z_0=\a_{2j}$, then  $h(z;x_b,t_b)=(z-z_0)^{m+\hf}[M +O(z-z_0)]$
in a vicinity of $z_0$, where $M\neq 0$ and $m=2,3,\cdots$. 
Therefore, the branchpoints $\a_{2k}$ satisfy the system
\be
\begin{cases}\label{modeg}
K(\a_{2k})&=0,~~~~k=0,1,\cdots, 2n, \cr
K^{({l})}(\a_{2j})&=0,~~~~ l=1,2,\cdots, m-1, \cr
\end{cases}
\ee
which is the system of modulation equations \eqref{modeq} for the branchpoints in the upper
halfplane with the requirement of additional degeneracy at $\a_{2j}$. According to 
\cite{TV2}, we can use $K(z)$ given by \eqref{K}.
As in part A), consider the subsystem  
\be
\begin{cases}\label{modegsub}
K(\a_{2k})&=0,~~~~k=0,1,\cdots,j-1,j+1,\cdots, 2n, \cr
K^{({m-1})}(\a_{2j})&=0, \cr
K(\a_{2j})&=0 \cr
\end{cases}
\ee
of \eqref{modeg}, 
which is a system of $2n+2$ complex equations for $2n+1$ complex variables $\a_{2k}, k=0,1,\cdots, 2n$,
and two real variables $x,t$. As it was shown in \cite{TV2}, the Jacobian matrix of the first $2n$
equations with respect to the variables $\a_{2k}, k=0,1,\cdots,j-1,j+1,\cdots, 2n$ is diagonal and 
invertible. Since $M\ne 0$, one can show that, similarly to \cite{TV2}, 
$\frac{\part}{\part \a_{2j}}K^{({m-1})}(\a_{2j})\ne 0$. So, in order to prove that the Jacobian
of \eqref{modegsub} is nonzero, it remains to show that 
\be\label{detKbarK}
\left| \begin{matrix}
K_x(\a_{2j}) & K_t(\a_{2j}) \cr
K_x(\bar \a_{2j})& K_t(\bar \a_{2j}) \cr
\end{matrix}
\right| \ne 0~,
\ee
where the fact that $K(z)$ is Schwarz symmetrical was taken into account.
Our arguments now are similar to those of part A). 
If vector $Q(\a_{2j})\ne 0$ (see \eqref{dKdxtgen}), then the rows of the latter determinant
are nonzero. Suppose $Q(\a_{2j})= 0$.  Consider the meromorphic differential
$\eta=\frac{d\z}{(\z-\a_{2j})R(\z)}$, whoose only pole is $\z=\a_{2j}$. This is an abelian 
differential of the second kind with zero periods. So, it satisfies \eqref{bilin2a}, where $\o$
is an arbitrary
abelian differntial, which cannot be true. Thus, the rows of \eqref{detKbarK} are nonzero.

To complete the proof, it is sufficient to show that for any $\x\in\C$ the vector 
\be\label{Vxz}
W(\x)= (W_1(\x),W_2(\x))=\left( K_x(\a_{2j})+\x K_x(\bar \a_{2j}),  
K_t(\a_{2j})+\x K_(\bar \a_{2j})\right)
\ee
is not zero. Components of $W(\x)$ can be represented as
\be\label{v12}
W_1(\x)=\frac{1}{D} \tilde K_1(\x),~~~~W_2(\x)=\frac{1}{D} \tilde K_2(\x),
\ee
where determinants $\tilde K_{1,2}(\x)$ are obtained from determinants
$\frac{\part}{\part x} K(z)$ and $\frac{\part}{\part t} K(z)$ in
\eqref{dKdxtgen} respectively by replacing
the last column $Q(z)$ with 
%\be\label{Zxz}
$Y(\x)=Q(\a_{2j})-\x Q(\bar \a_{2j})$.
%\ee
If  vector $Y(\x)\neq 0$ then
%, as in the proof  of Theorem \ref{hthxneq0}, we can establish that 
$W_1(\x)$ and $V_2(\x)$ cannot be zero simultaneously and the proof is completed.
In the remaining case $Y(\x)=0$ we consider the meromorphic differential
\be\label{etadiff}
\eta=\frac{d\z}{(\z-\a_{2j})R(\z)}+\x \frac{d\z}{(\z-\bar\a_{2j})R(\z)}
\ee
on $\Rscr$. This is an abelian differential of the second kind with poles at $\z=\a_{2j}$
and $\z=\bar \a_{2j}$. If vector $Y(\x)= 0$ then all the periods of $\eta$ are zero and,
using \eqref{bilin2a} as above, we obtain a contradiction.

%\begin{remark}\label{multiz0}
C) So far we considered only the case when at the breaking point $(x_b,t_b)$
the topology of zero level curves of $\Im h(z;x,t)$ in the spectral plane 
changes only at one point $z_0$. In general, it is possible that the change 
of topology occurs at two (or more) points $z_0$ and $z_1$ simultaneously
(note though that the same two branches of $\Im h(z;x,t)=0$ cannot 
intersect more than one time). Assuming that both $z_0$ and $z_1$ are double
points, we have two sets of equations \eqref{dpteq} valid at $z=z_0$ and $z=z_1$
with the same  $x=x_b, t=t_b)$. Thus we have six real equations for six real unknowns
which, according to \eqref{jac1}, have a nonvanishing Jacobian. Thus, such breaking points 
$(x_b,t_b)$ are isolated points on the $x,t$-plane.
The proof of the theorem is completed.
\ep

We now use  Theorem \ref{equalh}, as well as the results of this section, to prove the regular continuation principle in the case when all the branchpoints are bounded  
and stay away from the real axis.

\begin{theorem}\label{maintheo}
Let the nonlinear steepest descent asymptotics for solution $q(x,t,\e)$ of the NLS \eqref{NLS} be valid at some  point $(x_b,t_b)$. If $(x_*,t_*)$ is an arbitrary point, connected with $(x_b,t_b)$ by a
piecewise-smooth path $\Sigma$, if the countour 
$\g(x,t)$ of the RHP \eqref{rhpg} does not interact with  singularities of $f_0(z)$
as $(x,t)$ varies from $(x_b,t_b)$ to   $(x_*,t_*)$ along $\Sigma$, and if
all the branchpoints are bounded  
and stay away from the real axis,
%breaking points (if any) on $\Sigma$ are double points,
then  the nonlinear steepest descent asymptotics (with the proper choice of the  genus) is also valid at $(x_*,t_*)$.
\end{theorem}

\begin{figure}
%\begin{sideways}
\centerline{
\includegraphics[height=2.8cm]{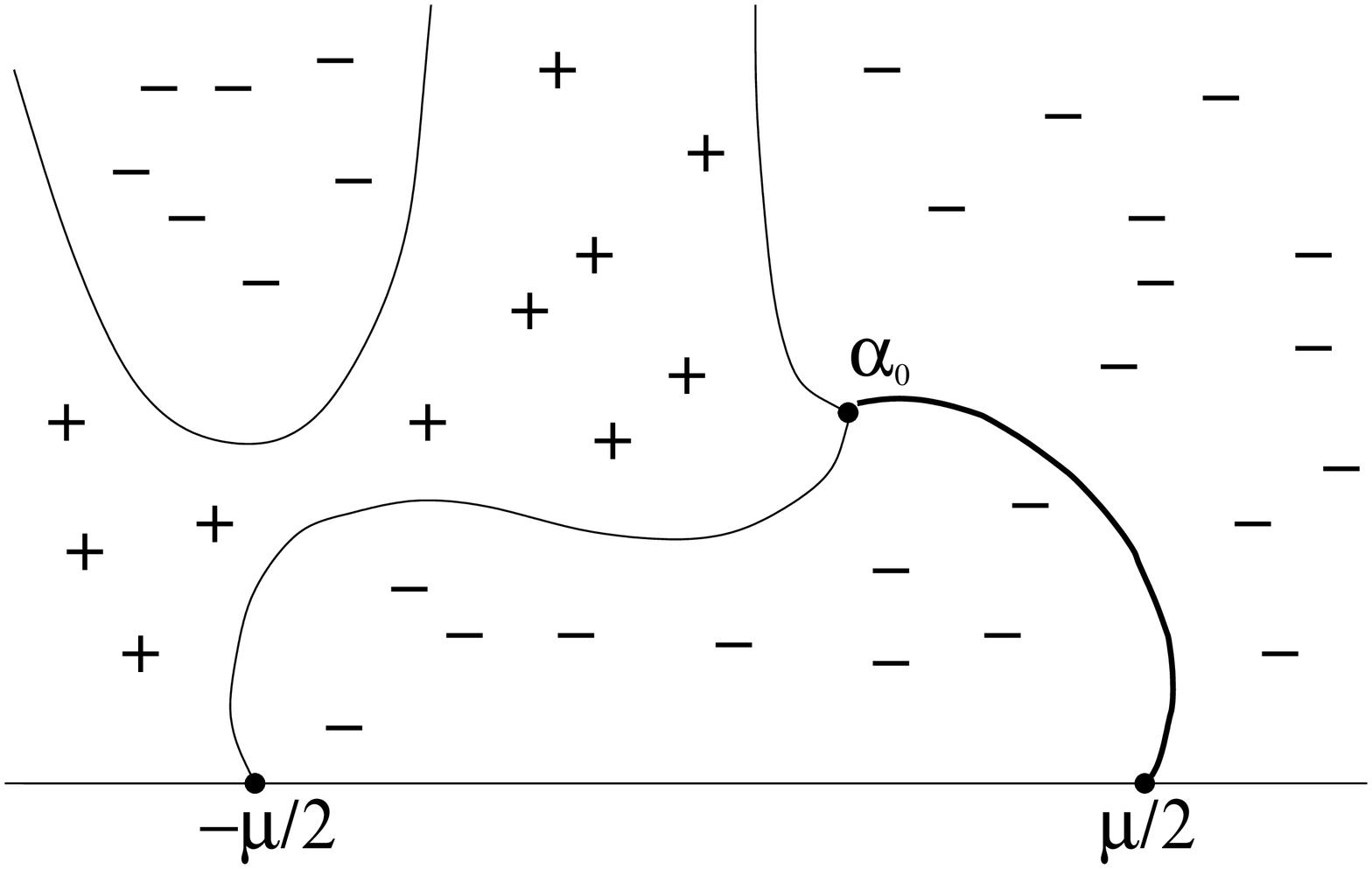}
\hskip 0.3cm
\includegraphics[height=2.8cm]{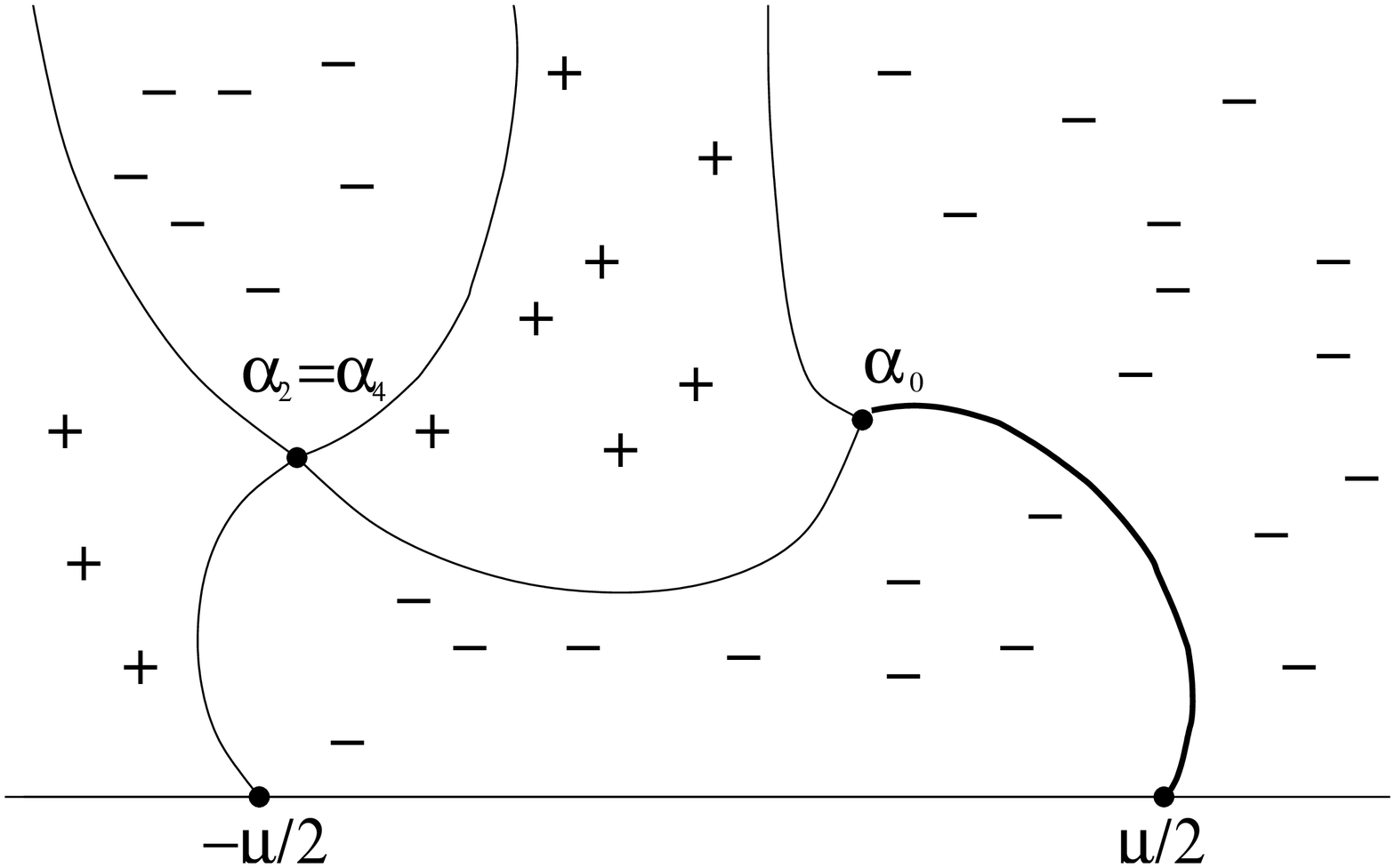}
\hskip 0.3cm
\includegraphics[height=2.8cm]{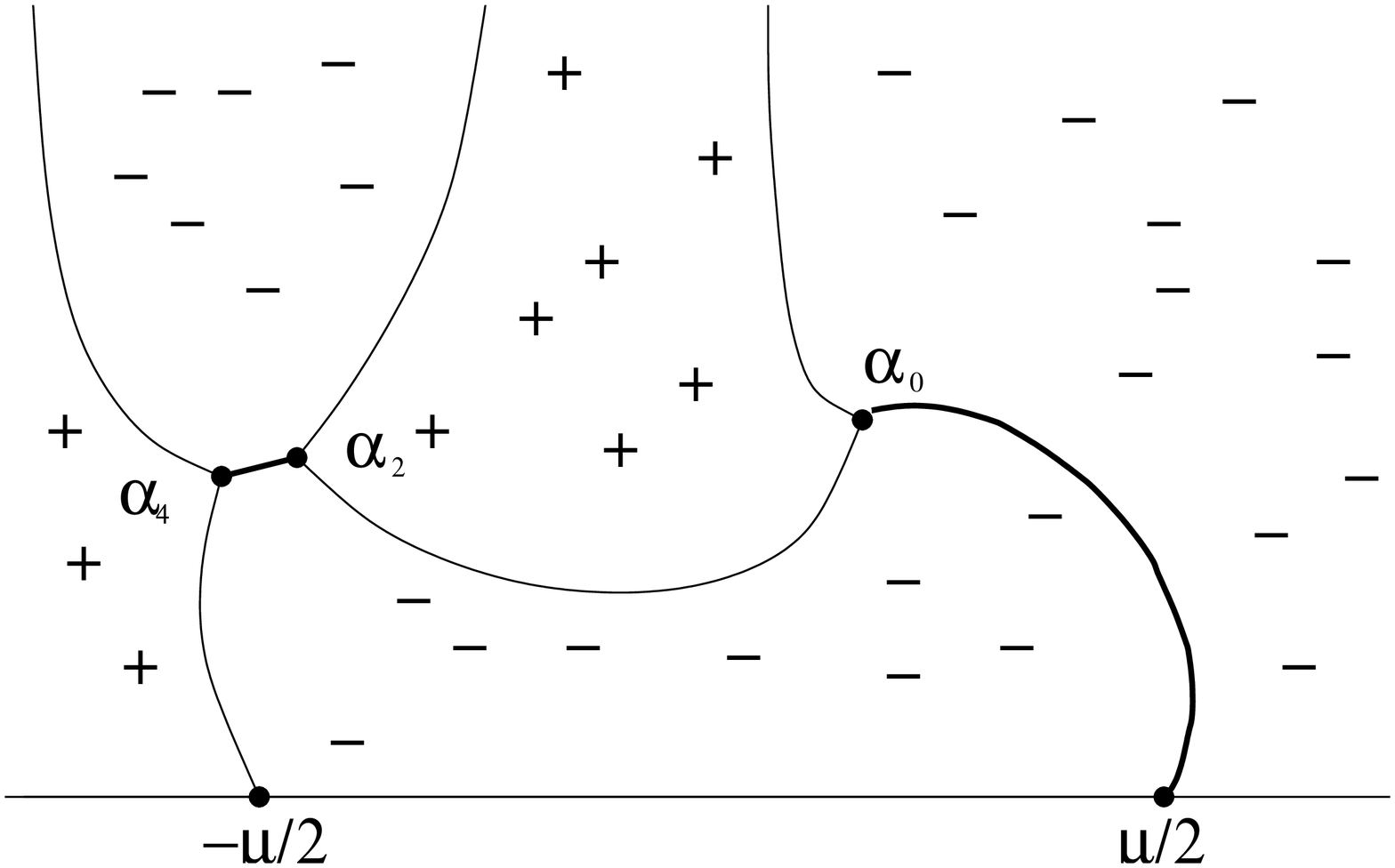}
}
%\vskip -.5truecm 
\caption{Transition from genus $N=0$ to genus $N=2$, where
\eqref{ineq} for the complementary arc $\g_{c}$ fails 
at $z_0\in \g_{c}$ (center). 
Zero level curves and signs of $\Im h$ are shown: left, before the break, $N=0$; center, at the break, $\Im h^(0)=\Im h^(2)$; right,
after the break, $N=2$. Note that $z_0=\a_2=\a_4$ at the break.}
%\hspace{.5cm}
%\psfig{file=../Figures/fig1b.ps}
%\includegraphics[height=9cm]{../Figures/fig1bl.ps}
%\end{center}
%\end{sideways}
\label{break}
\end{figure}

\bp
Let point $(x_b,t_b)$ belongs to the genus $N=2n$ region, $n\in\N$ of the solution $q(x,t,\e)$.
If $\Si$ does not intersect any breaking curve, or can be continuously deformed so that 
it does not intersect any breaking curve (while still satisfying the conditions of the theorem),
the proof follows from the Evolution Theorem of \cite{TVZ1}.
Otherwise, suppose traversing $\Si$  we find that at  some $(x_b,t_b)\in\Si$ (breaking point)
the inequalities \eqref{ineq} fail, say, 
at $z_0\in \g_{c,j}$. According to Theorem \ref{degisol}, we can assume that: $z_0$ is the only 
breaking point in the (upper) spectral plane corresponding to  $(x_b,t_b)$, and;
$z_0$ is a double (nondegenerate) breaking point. Otherwise $(x_b,t_b)$ is a degenerate 
breaking point that can be avoided by a small deformation of $\Si$.
Then, by Theorem \ref{brcurve}, there is a breaking curve $l$ passing through
$(x_b,t_b)$. If inequality \eqref{ineq} for the arc $\g_{c,j}$ fails only at one point $(x_b,t_b)$
of the contour $\Si$, i.e., if it  holds on $\Si$ on  a (punctured) vicinity of $(x_b,t_b)$,
then  the breaking point $(x_b,t_b)$  can be removed
by a small variation of $\Si$. Otherwise, we can assume that $\Si$ is transversal to $l$ at 
$(x_b,t_b)$. Then $D_\Si \Im h(z_0;x,t)|_{(x,t)=(x_b,t_b)}\le 0$ (see Fig. \ref{break}),
where $D_\Si$ denotes the directional derivative along $\Si$. Moreover,
according to Theorem
\ref{Imhthxneq0}, 

\be\label{Dirderold<0}
D_\Si \Im h(z_0;x,t)|_{(x,t)=(x_b,t_b)}<0. 
\ee

%Although the system of inequlities \eqref{ineq}, in general, is not valid past the point  $(x_b,t_b)\in\Si$, one can 
%try to change the genus of the problem from $N$ to $N+2$ hoping that the new system of inequalities
%\eqref{ineq} with the genus $N+2$ will be valid past the point  $(x_b,t_b)\in\Si$. 

%Assuming that $\a$ is a double point, i.e., that  $h_{zz}(z,x_b,t_b)|_{z=\a}\ne 0$,  one can  

Let us plant two additional branchpoints $\a_{4n+2}, \a_{4n+4}$  at $z_0$, $2n=N$,
which will open up a new main arc $\g_{m,n+1}$ as we move along $\Si$ past  the point $(x_b,t_b)$. 
According to \eqref{hform},  $h(z;x,t)=h^{(N)}(z;x,t)$ has a 
different expression in the genus $N+2$ region, 
i.e., beyond  the point $(x_b,t_b)\in\Si$, which we denote by $h^{(N+2)}$.
According to the Degeneracy Theorem from \cite{TVZ1},
\be\label{h=h} 
h^{(N+2)}(z;x_b,t_b)\equiv h^{(N)}(z;x_b,t_b)~.
\ee
%where $h^{(N)}$ denote ``old'' $h$.
The 
nonlinear steepest descend method asymptotics will remain 
valid on $\Si$ beyond  the point $(x_b,t_b)\in\Si$ if the ``newborn'' main arc $\g_{m,n+1}$ 
%with endpoints $\a_{4n+2}, \a_{4n+4}$ 
would also satisfy  \eqref{ineq}, that is, 
if $\Im h^{(N+2)}(z;x_b,t_b)<0$ to the left and to the right of $\g_{m,n+1}$.
The latter inequality will be satisfied if 
\be\label{Dirdernew<0}
D_\Si h^{(N+2)}(z;x,t)|_{(z;x,t)=(z_0;x_b,t_b)}<0. 
\ee
But \eqref{Dirdernew<0} follows from \eqref{Dirderold<0}, where $h(z;x,t)=h^{(N)}(z;x,t)$ and   
Theorem \ref{equalh}. Thus, the nonlinear steepst descent asymptotics with te genus $2N+2$ is valid
on $\Si$ beyond the breaking point $(x_b,t_b)$.    
The case when one of the  main arc inequalities of \eqref{ineq} is 
violated at $(x_b,t_b)$ can be treated similarly. 
The case when a main or a complementary arc collapses to a point can be treated as above
by moving in the opposite direction along $\Si$. So, we showed that the nonlinear steepest descent
asymptotics is valid ``automatically'' as a breaking curve is crossed, which implies the theorem.
\ep 

The authors are grateful to Marco Bertola for  insightful discussions related to abelian differentials.  

\bibliographystyle{plain}
\bibliography{spin.bib}

\begin{thebibliography}{10}

\bibitem{Deiftconf}
J.~Baik, Th. Kriecherbauer, and et~al.
\newblock {\em Integrable Systems and Random Matrices: In honor of Percy
  Deift}, volume 458.
\newblock AMS, 2008.

\bibitem{Belo}
E.~D. Belokolos, A.I. Bobenko, V.Z Enol'Skii, and A.R. Its.
\newblock {\em Algebro-geometric approach to nonlinear integrable equations}.
\newblock Springer, 1994.

\bibitem{DVZ2}
P.~Deift, S.~Venakides, and X.~Zhou.
\newblock New results in small dispersion kdv by an extension of the steepest
  descent method for riemann-hilbert problems.
\newblock {\em Internat. Math. Res. Notices}, 6:286--299, 1997.

\bibitem{DZ1}
P.~Deift and X.~Zhou.
\newblock A steepest descent method for oscillatory riemann - hilbert problems.
  asymptotics for the mkdv equation.
\newblock {\em Ann. of Math.}, 137:295--370, 1993.

\bibitem{DZ2}
P.~Deift and X.~Zhou.
\newblock Asymptotics for the painleve ii equation.
\newblock {\em Comm. Pure Appl. Math.}, 48(3):277--337, 1995.

\bibitem{KMM}
S.~Kamvissis, K.~T.-R. McLaughlin, and P.D. Miller.
\newblock Semiclassical soliton ensembles for the focusing nonlinear schr\"
  odinger equation.
\newblock {\em Annals of Mathematics Studies 154, Princeton Unversity Press},
  2003.

\bibitem{TV1}
A.~Tovbis and S.~Venakides.
\newblock Determinant form of modulation equations for the semiclassical
  focusing nonlinear schr�dinger equation.
\newblock {\em arXiv:0803.2066}, 2008.

\bibitem{TV2}
A.~Tovbis and S.~Venakides.
\newblock Determinant form of the complex phase function of the steepest
  descent analysis of riemann-hilbert problems and its application to the
  focusing nonlinear schr\" odinger equation.
\newblock {\em IMRN}, submitted.

\bibitem{TVZ1}
A.~Tovbis, S.~Venakides, and X.~Zhou.
\newblock On semiclassical (zero dispersion limit) solutions of the focusing
  nonlinear schroedinger equation.
\newblock {\em Comm. Pure Appl. Math}, 57(7):877--985, 2004.

\bibitem{TVZ3}
A.~Tovbis, S.~Venakides, and X.~Zhou.
\newblock Semiclassical focusing nonlinear schroedinger equation i: Inverse
  scattering map and its evolution for radiative initial data.
\newblock {\em International Mathematics Research Notices}, 2007(ID rnm094):54
  pages. doi:10., 2007.

\end{thebibliography}
\end{document}